\newcommand{\Hil}{{\mathcal H}}
\newcommand{\Ell}{{\mathcal L}}
\newcommand{\Prho}{{\mathcal P}}
\newcommand{\Boltz}{ k_{\rm\scriptscriptstyle B}}
\newcommand{\Tr}{{\rm Tr}}
\newcommand{\Ran}{{\rm Ran}}
\newcommand{\Ker}{{\rm Ker}}
\newcommand{\D}{{\bf \hat D}}
\newcommand{\F}{{\bf \hat F}}
\newcommand{\J}{{J\vphantom{\overline{J}}}}
\newcommand{\Jbar}{{\overline{J}}}
\newcommand{\ddt}[1]{{\frac{{\rm d}#1}{{\rm d}t}}}
\newcommand{\DDt}[1]{{\frac{{\rm D}#1}{{\rm D}t}}}
\newcommand{\sq}{{\sqrt{\rho}}}
\newcommand{\sqempty}{{{\scriptscriptstyle \sqrt{}}}}
\newcommand{\sqJ}{{\sqrt{\rho_\J}}}
\newcommand{\sqJbar}{{\sqrt{\rho_\Jbar}}}
\newcommand{\sqdotD}{{E_{D\vphantom{\overline{J}}}}}
\newcommand{\sqdotH}{{E_{H\vphantom{\overline{J}}}}}
\newcommand{\sqdotJD}{{E_{DJ\vphantom{\overline{J}}}}}
\newcommand{\sqdot}{{E_{D\vphantom{\overline{J}}}}}
\newcommand{\sqdotJ}{{E_{DJ\vphantom{\overline{J}}}}}

\newcommand{\pardsq}[1]{{\frac{{\partial}#1}{{\partial}\sqrt{\rho}}}}
\newcommand{\pardsqJ}[1]{{\frac{{\partial}#1}{{\partial}\sqrt{\rho_\J}}}}
\newcommand{\hpardsq}[1]{{ \partial #1/\partial\sqrt{\rho}}}
\newcommand{\pardsqdot}[1]{{\frac{{\partial}#1}{{\partial}\sqdot }}}

\newcommand{\cov}[2]{{\langle\Delta #1\Delta #2\rangle}}
\newcommand{\Otimes}{{ \otimes }}

\documentstyle[pra,aps,twocolumn]{revtex}

\begin{document}

\title{Maximal-entropy-production-rate nonlinear quantum dynamics compatible with second law, reciprocity, fluctuation--dissipation, and time--energy uncertainty relations}
\author{ Gian Paolo Beretta }
\address{
Istituto Nazionale di Fisica della Materia\\
Universit\`a di Brescia, via Branze 38, 25123 Brescia, Italy\\
{\tt beretta@ing.unibs.it}\\
{\rm (submitted to Phys.Rev.A on February 9, 2001; revised September 14, 2001)}\\
{\rm (e-print: quant-ph/0112046)}
}

\maketitle
\begin{abstract}
In view of the recent quest for well-behaved nonlinear extensions of the traditional Schr\"odinger--von Neumann unitary dynamics that could provide fundamental explanations of recent experimental evidence of loss of quantum coherence at the microscopic level, I review the general features and complete with an additional fundamental ansatz the nonlinear quantum (thermo)dynamics I proposed [in my doctoral thesis (MIT, 1981) and the subsequent series of papers cited in G.\,P.\ Beretta, Found.\ Phys.\ {\bf 17}, 365 (1987)] for a single isolated indivisible constituent system, such as a single particle or spin or atomic system, or a Bose-Einstein or Fermi-Dirac field.
The nonlinear dynamics entails a fundamental unifying microscopic proof and extension of Onsager's reciprocity  and Callen's fluctuation--dissipation relations  to all nonequilibrium states, close and far from thermodynamic equilibrium. Together with a self-contained review of the main results already proved, I show an explicit geometrical construction of the equation of motion from the steepest-entropy-ascent ansatz and clarify its exact mathematical and conceptual equivalence with the maximal-entropy-production variational-principle formulation presented in S.\ Gheorghiu-Svirschevski, Phys.\ Rev.\ A {\bf 63}, 022105 (2001), and I show how it can be extended to the case of a composite system to obtain the general form of my equation of motion, consistently with the demanding requirements of strong separability.
Because both my original geometric construction and its equivalent variational formulation, fix only the direction (of steepest entropy ascent) along which the state operator is driven by the dissipative term of the equation of motion, but leave unspecified the rate at which this happens, I propose a possible, well-behaved and intriguing, general closure of the dynamics, compatible with the nontrivial requirements of strong separability.
Based on the time--energy Heisenberg uncertainty relation, I derive lower bounds to the internal-relaxation-time functionals that determine the rate of entropy production. This bounds entail an upper bound to the rate of entropy production.  Guided by this reasoning, I propose a maximal-entropy-production-rate ansatz, by which each indivisible subsystem follows the direction of steepest perceived entropy ascent at the highest rate compatible with the time--energy uncertainty principle. Subject, of course, to experimental and further theoretical validation, this closure ansatz completes the original nonlinear dynamics with no need of new physical constants.

\end{abstract}


\narrowtext

\section{Introduction}

In my 1981 doctoral thesis \cite{thesis} and a subsequent series of papers \cite{Cimento,Beretta,Attractor,Fluorescence,Onsager,ASME,Taormina} (see also Refs.\ \cite{Korsch,Janussis}), I addressed the problem of deriving a well-behaved extension of the Schr\"odinger--von Neumann unitary dynamics meeting the very demanding set of strict requirements that appear to be necessary if one is willing to accept \cite{HG}: (1) the ``broader quantum kinematics'' ansatz that $\rho$ --- a unit-trace, nonnegative-definite, hermitian operator on the Hilbert space $\Hil$ associated with a system according to standard quantum mechanics --- represents a ``true'' quantum state (in the sense analogous to that of the wave function of standard QM) not only when $\rho^2=\rho$ but also when $\rho^2\ne\rho$, even if the system is strictly isolated and uncorrelated; and (2) that $-\Boltz \Tr (\rho \ln \rho)$ represents the ``physical'' entropy (as opposed to a statistical or information-theoretic entropy related to incoherent stochastic mixtures of true states).

The equation of motion I designed and proposed for the state operator $\rho$ of a general system consisting of $M$ distinguishable  indivisible subsystems has the form
\begin{mathletters}\label{EquationComposite}
\begin{eqnarray}&&\qquad \qquad {\displaystyle \ddt{\rho} = - \frac{i}{\hbar}}[H,\rho] +\D_M (\rho,H,G_i ) \ , \\
&&\D_M ( \rho,H,G_i ) = -{\displaystyle \sum_{J=1}^M \frac{1}{ 2\tau_\J(\rho)} }\big[\sqJ D_\J\!+\!D_J^\dagger\sqJ\big]\Otimes \rho_{\overline{J}} \ , \nonumber \\ \\
&&\  D_\J= [\sqJ (B\ln\rho)^\J]_{\bot\Ell {\{\sqJ I_\J,\sqJ (H)^\J, [\sqJ (G_{i })^J]\}}} \ ,
\end{eqnarray}
\end{mathletters}
where the notation is defined in Section \ref{Composite}.

For a single indivisible system, it reduces to the form
\begin{mathletters}\label{EquationSingle}
\begin{eqnarray}
&{\displaystyle \ddt{\rho} = - \frac{i}{\hbar}} [H,\rho] +\D_1 (\rho,H,G_i )\ ,& \\
&\D_1 (\rho,H,G_i )= -{\displaystyle \frac{1}{2\tau (\rho)}} \big[\sq D +D ^\dagger\sq\big] \ ,&\\
&D = [\sq\ln\rho]_{\bot\Ell {\{\sq I,\sq H, [\sq G_i ]\}}} \ ,&
\end{eqnarray}
\end{mathletters}
where $[\sq\ln\rho]_{\bot\Ell {\{\sq R_{i } \}}}$ denotes the component of operator $\sq\ln\rho$ orthogonal to the linear manifold $\Ell {\{\sq R_{i } \}}= \Ell \{\sq I, \sq H, [\sq G_i]\}$ spanned by all real linear combinations of the set of operators $\{\sq I, \sq H, [\sq G_i]\}$ where $I$ is the identity operator on the Hilbert space $\Hil$ of the system, $H$ the usual Hamiltonian operator associated with the system in standard QM, the $G_i$'s are additional (not always necessary) hermitian operators commuting with $H$ that we call the non-Hamiltonian generators of the motion and that depend on the structure of the system, such as the $i$-th-type particle-number operators $N_i$ for a Bose-Einstein or Fermi-Dirac field (in which case $\Hil$ is a Fock space), or the total momentum component operators $P_i$ of a free particle (in which case Gheorghiu-Svirschevski \cite{Gheorghiu} proved Galilei invariance).

As I realized in Ref.\ \cite{Onsager}, the internal-relaxation times $\tau_\J (\rho)$ need not be constants (as initially assumed in Ref.\ \cite{Cimento}) but can be any positive functionals of $\rho$.  However, here I note that the condition of separability imposes precise restrictions on these functionals. In Section \ref{Heisenberg}, I derive a lower bound for $\tau_\J (\rho)$ which corresponds to the highest entropy production rate compatible with the time--energy Heisenberg uncertainty principle.  Taking the internal-relaxation times $\tau_\J (\rho)$ exactly equal to these nontrivial lower bounds turns out to be compatible with the separability condition, and completes the nonlinear dynamics in a general way with no need to assume the existence of new physical constants.

Equation (\ref{EquationComposite}) satisfies all the necessary
requirements (listed in Appendix \ref{criteria}) for a
self-consistent and well-defined extension of the
Schr\"odinger--von Neumann equation of motion compatible with thermodynamics requirements, and turns out to
have very intriguing mathematical properties and consistent
physical consequences.

In this paper, in addition to providing a self-contained review of results already proved in previous work, I give a more explicit account of the steepest-entropy-ascent geometric construction on my nonlinear equation of motion for quantum thermodynamics than available in the original papers.  As a result, I show that the equivalent variational formulation in Ref.\ \cite{Gheorghiu} can be extended to the composite system case.  I also show that, in order for the general form of my equation to satisfy strong separability conditions conceptually equivalent to those recently discussed by Czachor \cite{Czachor}, the relaxation-time functionals must satisfy nontrivial conditions.

I emphasize that the nonlinearity of the dynamics is such that one should not expect that the equation valid for a single elementary (indivisible) constituent of matter be valid also for a system with an internal structure, because the equation of motion must reflect such structure not only through the Hamiltonian operator.  However, of course, the superoperator $\D_M$ must reduce to $\D_1$ [see Eq.\ (\ref{rhodotM}) below] when the system has only one indivisible subsystem.

For a composite system, my original general formulation
To make the paper self-contained, I briefly review the fundamental unifying extensions to all nonequilibrium (dissipative) states of the Onsager reciprocity relations \cite{Lars} and Callen's fluctuation-dissipation theorem \cite{Callen} that derive from Eq.\ (\ref{EquationSingle}) \cite{Onsager} as well as, in general, from Eq.\ (\ref{EquationComposite}) \cite{Taormina}.

In Appendixes \ref{criteria} and \ref{stability}, I discuss a set of criteria and the definition of stability of equilibrium that a fundamental dynamics must satisfy in order to be compatible with the second law. In the other Appendixes, I discuss some extensions of the mathematics of Eqs.\ (\ref{EquationComposite}) and (\ref{EquationSingle}).

\section{The augmented state domain ansatz}\label{Conceptual}

The fundamental ansatz  that the postulates of quantum mechanics can be successfully supplemented by the first and second principles of thermodynamics by assuming a broader state domain that includes not only $\rho^2=\rho$ but also $\rho^2\ne\rho$ state operators, provided that the functional $-\Boltz\Tr(\rho\ln\rho)$ is taken for the physical entropy, was first proposed (without a dynamical law) by Hatsopoulos and Gyftopoulos \cite{HG}.

Twenty five years ago, the hypothesis of a state domain augmented with respect to that of traditional QM was perceived as countercurrent to the prevailing approaches to dissipative quantum dynamics within the frameworks of statistical, stochastic, phenomenological, information-theoretic, chaotic-behavior and bifurcation theories.  For this reason, the broader quantum kinematics ansatz \cite{HG} and, five years later, my equation of motion have been considered initially as unphysical and substantially ignored, mainly because their motivation appeared to be derived from theoretical reasoning only.

In search for direct experimental evidence, I computed the effects of the irreversible atomic relaxation implied by the nonlinear equation of motion onto some basic quantum-electrodynamics results on absorption, stimulated emission, and resonance fluorescence from a single two-level atom \cite{Attractor,Fluorescence}. The results were obtained in the near-equilibrium linear limit and, of course, in terms of the yet undetermined internal-relaxation-time functional $\tau(\rho)$ that is part of the equation of motion. To my knowledge no one has yet attempted to verify these results experimentally and estimate $\tau(\rho)$.

The recent new experimental evidence of loss of quantum coherence \cite{Experimental,Reznik} and the impressive effort devoted to study nonlinear modifications of the standard Schr\"odinger equation in the last ten years \cite{nonlinearS}, finally seem to make more acceptable, if not require, the $\rho^2\ne\rho$ augmented state domain Hatsopoulos--Gyftopoulos ansatz.

Once the $\rho^2\ne\rho$ ansatz is accepted, the nonlinear equation of
motion I proposed completes the dynamics and holds the promise to
provide a microscopic-level explanation of the recent experimental
evidence of loss of quantum coherence. It is with this motivation that
Gheorghiu-Svirschevski \cite{Gheorghiu} has ``rediscovered'' the
simplest form of my equation together with many of its known features.
Ref.\ \cite{Gheorghiu} contributes to confirm the validity of my
equation, including existence and uniqueness of solutions, and derives
useful results in the near-equilibrium linear limit.  However, the
question of defining the form of the relaxation-time functional is left
unresolved.

\section{The steepest-entropy-ascent ansatz for an indivisible system}\label{Steepest}

In Refs.\ \cite{Beretta,Onsager,ASME}, I emphasized that Eqs.\ (\ref{EquationComposite}) and (\ref{EquationSingle}) have an important geometric interpretation.  The Hamiltonian term, $-i[H,\rho]/\hbar$, and the dissipative term, $\D (\rho,H,G_i)$, compete with each other in the sense that the first drives the state operator toward a unitary motion tangent to the local constant entropy surface, whereas the second drives it toward the local direction of steepest entropy ascent along the surface with constant mean values of the generators of the motion.  In this section, I show more explicitly than in the original papers how Eq.\ (\ref{EquationSingle}) can be constructed directly from the steepest-entropy-ascent ansatz.

Let $\Hil$ (${\rm dim}\Hil \le \infty$) be the Hilbert space and $H$ the Hamiltonian operator that are associated with the given (indivisible) elementary constituent system in standard QM.  For simplicity, we first consider a system composed of a single elementary constituent.  The generalization to $M$ constituents is given in Section \ref{Composite}.

We assume that the (true) quantum states are one-to-one with the linear hermitian operators $\rho$ on $\Hil$ with $\Tr(\rho)=1$ and $\rho\ge 0$.  As done in Ref.\ \cite{Cimento}, we introduce the square-root state operator $\sq$ obtained from $\rho$ by substituting its eigenvalues with their positive square roots \cite{NotaSquareRoot}.

As a first step to force positivity and hermiticity of the state
operator $\rho$ we assume that the equation of motion may be written as
\begin{equation}
\label{rhodot}\ddt{\rho}=\sq\,E+ E^\dagger\!\sq
\end{equation}
where the operator $E$, in general non-hermitian, is defined later.

We consider the space $\Ell (\Hil)$ of linear operators on $\Hil$ equipped with the real scalar product

\begin{equation}
\label{Scalar} (F|G) = \frac{1}{2}\Tr (F^\dagger G + G^\dagger F)  \ .
\end{equation}
so that for any (time-independent) hermitian $R$ in $\Ell (\Hil)$ the corresponding mean-value state functional, its local gradient operator with respect to $\sq$, and its rate of change are
\begin{eqnarray}
\label{meanR} r(\rho)&=&\Tr (\rho R)=(\sq|\sq R) \ ,\\ \label{gradR}
\nabla_{\!\sq\,}r(\rho)&=&\pardsq{r(\rho)}= \sq R+ R\sq\ ,\\
\label{rateR} \ddt{r(\rho)}&=&\Tr(
\ddt{\rho}R)=2\left.\left(E\right|\sq R\right) \ .
\end{eqnarray}

Moreover, the entropy state functional, its gradient operator with respect to $\sq$, and its rate of change are
\begin{eqnarray}
\label{meanS} s(\rho)&=&-\Boltz\Tr (\rho\ln\rho)=-\Boltz (\sq |\sq
\ln\rho) \ ,\\ \label{gradS}\nabla_{\!\sq \,}s(\rho)&=&
\pardsq{s(\rho)}=-2\Boltz\left[\sq +\sq\ln\rho\right]\ ,\\
\label{rateS} \ddt{s(\rho)}&=& -\Boltz \left[\Tr(\ddt{\rho})+
\Tr(\ddt{\rho}\ln\rho) \right]\nonumber \\ &=&
\left(E\left|\pardsq{s(\rho)}\right.\right)=
\left(\left.\pardsq{s(\rho)} \right|E\right) \ .
\end{eqnarray}

Now it is easy to see from Eq.\ (\ref{rateR}) that the values of the
mean functionals $r_i(\rho)$ are time invariant if and only if E is
orthogonal to $\sq R_i$, for all $i$, i.e., if it is orthogonal to the
linear manifold $\Ell {\{\sq R_{i } \}}$  spanned by the set of
operators $\{\sq R_{i } \}$ in which we always have $\sq R_0=\sq I$ (to
preserve $\Tr \rho=1$) and $\sq R_1=\sq H$ (to conserve energy), plus
the additional non-Hamiltonian generators of the motion as already
discussed.

It is noteworthy that
\begin{equation}\label{sqHamil}
\sqdotH=\frac{i}{\hbar}\sq\, [H+c(\rho)I] \ ,
\end{equation}
with $c(\rho)$ any real functional of $\rho$, yields [through Eq.\ (\ref{rhodot})] the Hamiltonian part of the equation of motion, and is orthogonal to $\Ell {\{\sq R_{i } \}}$ and to the entropy gradient $\hpardsq{s(\rho)}$ as well.

Introducing the notation \cite{Onsager}
\begin{eqnarray}
\label{Delta}\Delta F&=&F- \Tr(\rho F)I \ ,\\
\label{Cov}\cov{F}{G}&=& \cov{G}{F} =(\sq \Delta F|\sq \Delta G)
\nonumber\\ &=&\frac{1}{2} \Tr(\rho\{\Delta F,\Delta G\})\ ,
\end{eqnarray}
for $F$ and $G$ hermitian in $\Ell (\Hil)$ and $\{\cdot,\cdot\}$ the usual anticommutator, and
defining the shortest characteristic time associated with the Hamiltonian part of the equation of motion by the relation (see Appendix \ref{times})
\begin{equation}\label{normH1}
\frac{1}{\tau_H(\rho)^2}=4\left(\left.\sqdotH\right|\sqdotH\right) \ ,
\end{equation}
we find, from Eq.\ (\ref{sqHamil}) and $(\sq|\sq)=1$,
\begin{equation}\label{normH2}
\tau_H(\rho)^2 \cov{H}{H}=\hbar^2/4-[c(\rho)+\Tr(\rho
H)]^2\tau_H(\rho)^2 \ ,
\end{equation}
so that only the choice $c=-e(\rho)=-\Tr (\rho H)$ is compatible with the Heisenberg time--energy uncertainty relation $\tau_H(\rho)^2 \cov{H}{H}\ge\hbar^2/4$ which is then satisfied with strict equality.  Therefore,
\begin{equation}\label{sqHamilDelta}
\sqdotH=\frac{i}{\hbar}\sq\, \Delta H \ ,
\end{equation}
and
\begin{equation}\label{normH3}
\tau_H(\rho)^2 \cov{H}{H}=\hbar^2/4 \ .
\end{equation}

We may note that the foregoing discussion on the choice of $c(\rho)$ is somewhat artificial because any $c(\rho)$ cancels out in Eq.\ (\ref{rhodot}).  However, the line of thought becomes important when we apply it later to the dissipative part of the dynamics.

For the general rate of change of $\sq$, we let
\begin{equation}
\label{sqdot}E =\sqdotH + \sqdotD \ ,
\end{equation}
and assume $ \sqdotD $ orthogonal to $\Ell {\{\sq R_{i } \}}$ but in the direction of the entropy gradient operator.  We cannot take $ \sqdotD $ directly proportional to $\hpardsq{s(\rho)}$ as such, because the entropy gradient in general has a component along $\Ell {\{\sq R_{i } \}}$ which would not preserve the mean values of the generators of the motion. We must take $ \sqdotD $ proportional to the component of the entropy gradient orthogonal to $\Ell {\{\sq R_{i } \}}$, namely, for the indivisible system,
\begin{mathletters}\label{sqdotD}\begin{eqnarray}
\sqdotD  &&=\frac{1}{4\Boltz \tau(\rho)}\left[\pardsq{s(\rho)}\right]_{\bot\Ell {\{\sq R_{i } \}}} \\&&=\frac{1}{4\Boltz\tau(\rho) }\left(\pardsq{s(\rho)}-\left[\pardsq{s(\rho)}\right]_{ \Ell {\{\sq R_{i } \}}}\right) \\
&&=-\frac{1}{2\tau(\rho) }\left(\sq\ln\rho-\left[\sq\ln\rho \right]_{ \Ell {\{\sq R_{i } \}}}\right) \\
&&=-\frac{1}{2\tau(\rho) } D \label{rhodotD}
\end{eqnarray}\end{mathletters}
where $[\hpardsq{s(\rho)}]_{ \Ell {\{\sq R_{i } \}}}$ denotes the projection of $ \hpardsq{s(\rho)}$ onto $ \Ell {\{\sq R_{i } \}}$ and
\begin{mathletters}\label{Dgram}\begin{eqnarray}
D&&= \sq\ln\rho-\left[\sq\ln\rho \right]_{ \Ell {\{\sq R_{i } \}}}\\ &&=\left[\sq\ln\rho \right]_{\bot \Ell {\{\sq R_{i } \}}} \\ \nonumber & \ &\\
&&=\frac{\left|
\begin{array}{ccccc}
\sq\ln\rho & \sq R_0 &\!\cdots\!& \sq R_i &\!\cdots\!\\ \\
({\scriptstyle \sq\ln\rho }|{\scriptstyle \sq R_0}) & ({\scriptstyle \sq R_0}|{\scriptstyle \sq R_0}) &\!\cdots\!& ({\scriptstyle \sq R_i}|{\scriptstyle \sq R_0}) & \!\cdots\!\\
\vdots & \vdots & \!\ddots\! & \vdots&\!\ddots\!\\
({\scriptstyle \sq\ln\rho} | {\scriptstyle \sq R_i}) & ({\scriptstyle \sq R_0}|{\scriptstyle \sq R_i}) &\!\cdots\!& ({\scriptstyle \sq R_i}|{\scriptstyle \sq R_i}) & \!\cdots\!\\
\vdots& \vdots & \!\ddots\! & \vdots&\!\ddots\!
\end{array}
\right|}
{\Gamma(\{\sq R_i\}) }
\nonumber\\ &\ &\\
&&=-\frac{1}{\Boltz}\frac{\left|
\begin{array}{ccccc}
\sq \Delta S & \sq \Delta R_1 &\!\cdots\!& \sq \Delta R_i &\!\cdots\!\\ \\
{\scriptstyle\cov{S}{R_1}}& {\scriptstyle\cov{R_1}{R_1}} &\!\cdots\!& {\scriptstyle\cov{R_i}{R_1}} & \!\cdots\!\\
\vdots & \vdots & \!\ddots\! & \vdots&\!\ddots\!\\
{\scriptstyle\cov{S}{R_i}}& {\scriptstyle\cov{R_1}{R_i}} &\!\cdots\!& {\scriptstyle\cov{R_i}{R_i}} & \!\cdots\!\\
\vdots& \vdots & \!\ddots\! & \vdots&\!\ddots\!
\end{array}
\right|}
{\Gamma(\{\sq R_i\}) }
\nonumber\\
&\ &
\end{eqnarray}\end{mathletters}
where, for convenience, we defined the entropy operator
\begin{mathletters}\begin{equation}
\label{Sdef} S=-\Boltz B\ln\rho
\end{equation}
with $B$ obtained from $\rho$ by substituting its nonzero eigenvalues with unity, i.e., more formally, $B=B^2=P_{\Ran\, \rho}=I- P_{\Ker\, \rho}= P_{\bot\Ker\, \rho}$, so that
\begin{equation}
\label{Bdef} B=B^2\ ,\quad [B,\rho]=0\ ,\quad B\rho=\rho\ .
\end{equation}\end{mathletters}

The Gram determinant at the denominator,
\begin{mathletters}\begin{eqnarray}
\Gamma(\{\sq R_i\})&=&\det[\{(\sq R_i|\sq R_j)\}]  \\
&=&\left|
\begin{array}{cccc}
 ({\scriptstyle \sq R_0}|{\scriptstyle \sq R_0}) &\!\cdots\!& ({\scriptstyle \sq R_i}|{\scriptstyle \sq R_0}) &\!\cdots\!\\
\vdots & \!\ddots \!& \vdots&\!\ddots\!\\
({\scriptstyle \sq R_0}|{\scriptstyle \sq R_i}) &\!\cdots\!& ({\scriptstyle \sq R_i}|{\scriptstyle \sq R_i}) &\!\cdots\!\\
\vdots & \!\ddots \!& \vdots&\!\ddots\!
\end{array}
\right|  \\
&=&\det[\{\cov{R_{i>0}}{R_{j>0}}\}] \\
&=&\left|
\begin{array}{cccc}
{\scriptstyle\cov{R_1}{R_1}} &\!\cdots\!& {\scriptstyle\cov{R_i}{R_1}} & \!\cdots\!\\
\vdots & \!\ddots\! & \vdots&\!\ddots\!\\
{\scriptstyle\cov{R_1}{R_i}} &\!\cdots\!& {\scriptstyle\cov{R_i}{R_i}} & \!\cdots\!\\
\vdots & \!\ddots\! & \vdots&\!\ddots\!
\end{array}
\right| \ ,
 \end{eqnarray}\end{mathletters}
is always strictly positive by virtue of the linear independence of the operators in the set ${\{\sq R_{i } \}}$.

A further, compact expression can be written if we choose a set of operators $\{\sq A_{i } \}$ which like the set
$\{\sq R_{i } \}$ spans $\Ell \{\sq I, \sq H, [\sq G_i]\}$ but, in addition, forms an orthonormal set, e.g., obtained from $\sq I$, $\sq H$, $\sq G_i$ by a Gram-Schmidt orthogonalization procedure followed by normalization. For example,
\begin{mathletters}\begin{eqnarray}
&A_1=I\ ,& \\
&A_2=\frac{\Delta H}{\sqrt{\cov{H}{H}}}\ , &\\
&A_3=\frac{\cov{H}{H}\Delta G_1-\cov{H}{G_1}\Delta H}{\sqrt{\cov{H}{H}[\cov{H}{H}\cov{ G_1}{ G_1}-\cov{H}{ G_1}^2]}}\ ,&\\
&\cdots\quad .&\nonumber
\end{eqnarray}\end{mathletters}
Then,
\begin{mathletters}\label{Orthonormal}
\begin{eqnarray}
&(\sq A_i|\sq A_j)=\delta_{ij}\ ,&\\
& D= \sq\ln\rho- \sum_{i=1}^a (\sq\ln\rho |\sq A_i) \sq A_i \label{DA}\ ,&\\
&\Gamma(\{\sq A_i\})=1 \ . &
\end{eqnarray}
\end{mathletters}
It is noteworthy that operators $A_i$ are nonlinear functions of $\rho$ and, therefore, in general, vary with $\rho$ as it evolves with time.

We finally obtain the following equivalent expressions for the rate of entropy change
\begin{mathletters}\label{rateSfinal} \begin{eqnarray}
\ddt{s(\rho)}&=& \frac{1}{4\Boltz\tau(\rho)}
\left(\!\left[\pardsq{s(\rho)}\right]_{\!\bot\Ell {\{\sq R_{i }
\}}}\!\left|\left[\pardsq{s(\rho)}\right]_{\!\bot\Ell {\{\sq R_{i }
\}}}\!\right) \right.\nonumber\\ && \\ &=&
{4\Boltz\tau(\rho)}\left(E\left|E\right.\right) \\
&=&{4\Boltz\tau(\rho)}\left(\sqdotD \left|\sqdotD \right.\right) \\
&=&\frac{1}{\Boltz\tau(\rho)}\left[\cov{S}{S} - \sum_{i=1}^a
(\cov{S}{A_i})^2\right] \\ &=& \frac{\Boltz }{ \tau(\rho)} (D|D)
\\&=& \frac{\Boltz }{ \tau(\rho)} \frac{\Gamma(\sq\ln\rho,\{\sq R_i\})}{\Gamma(\{\sq R_i\})}
\\&=& \frac{1 }{ \Boltz\tau(\rho)} \frac{\Gamma(\sq S,\{\sq R_i\})}{\Gamma(\{\sq R_i\})}
\end{eqnarray}\end{mathletters}
where the Gram determinant
\begin{eqnarray}
&&\Gamma(\sq S,\{\sq R_i\}) \nonumber\\
&&\quad=\left|
\begin{array}{ccccc}
({\scriptstyle \sq S }|{\scriptstyle \sq S\hphantom{_i} }) & ({\scriptstyle \sq R_0}|{\scriptstyle \sq S\hphantom{_i} }) &\!\cdots\!& ({\scriptstyle \sq R_i}|{\scriptstyle \sq S\hphantom{_i} }) & \!\cdots\!\\ \\
({\scriptstyle \sq S }|{\scriptstyle \,\sq R_0}) & ({\scriptstyle \sq R_0}|{\scriptstyle \,\sq R_0}) &\!\cdots\!& ({\scriptstyle \sq R_i}|{\scriptstyle \,\sq R_0}) & \!\cdots\!\\
\vdots & \vdots & \!\ddots\! & \vdots&\!\ddots\!\\
({\scriptstyle \sq S} | {\scriptstyle \,\sq R_i}) & ({\scriptstyle \sq R_0}|{\scriptstyle \,\sq R_i}) &\!\cdots\!& ({\scriptstyle \sq R_i}|{\scriptstyle \,\sq R_i}) & \!\cdots\!\\
\vdots& \vdots & \!\ddots\! & \vdots&\!\ddots\!
\end{array}
\right| \, \nonumber\\ \nonumber \\
&&\quad=\left|
\begin{array}{ccccc}
{\scriptstyle\cov{S}{S}} &{\scriptstyle\cov{R_1}{S}} &\!\cdots\!& {\scriptstyle\cov{R_i}{S}} & \!\cdots\!\\ \nonumber\\
{\scriptstyle\cov{S}{R_1}} &{\scriptstyle\cov{R_1}{R_1}} &\!\cdots\!& {\scriptstyle\cov{R_i}{R_1}} & \!\cdots\!\\
\vdots &\vdots & \!\ddots\! & \vdots&\!\ddots\!\\
{\scriptstyle\cov{S}{R_i}} &{\scriptstyle\cov{R_1}{R_i}} &\!\cdots\!& {\scriptstyle\cov{R_i}{R_i}} & \!\cdots\!\\
\vdots &\vdots & \!\ddots\! & \vdots&\!\ddots\!
\end{array}
\right| \\
\end{eqnarray}
is also strictly positive except when $\rho$ satisfies Eq.\ (\ref{Nondissipative}).

\section{Schr\"odinger--von Neumann limit}

Equation (\ref{EquationSingle}) reduces to the Schr\"odinger--von Neumann equation of motion, $i\hbar\dot\rho=[H,\rho]$, when $\rho^2=\rho$ (thus entailing the usual unitary hamiltonian dynamics of standard QM), and also when and only when
\begin{mathletters}\label{Nondissipative}
\begin{eqnarray}
\rho &=& \frac{B \exp(C)}{\Tr[ B \exp(C)]}\ ,\\
B^2&=& B\ , \quad [B,C]=0 \ ,
\end{eqnarray}
or, equivalently,
\begin{equation}
\label{Cnondiss}\sq\ln\rho = \sq C -\sq\ln\Tr[B\exp(C )]\ ,
\end{equation}
where
\begin{equation}
C= -\beta H+\sum_i\nu_i G_i \ ,
\end{equation}
\end{mathletters}
in which case we say that the state is nondissipative, and the solution of the equation of motion is
\begin{mathletters}\label{Unitary}
\begin{eqnarray}
&\rho(t) = B(t) \exp(C)/\Tr[B(t) \exp(C)]\ ,&\\
\label{Bunitary}&B(t) = U(t) B(0) U^{-1}(t) \ ,\quad  U(t)=\exp(-itH /\hbar)\ ,&
\end{eqnarray}
\end{mathletters}
which includes the usual $\rho^2=\rho$ Schr\"odinger dynamics when $C$ is the null operator [$\beta= \nu_i=0$ in Eq.\ (\ref{Nondissipative})] and, therefore, $\Tr (B)=1$.

When $[B(0),H]= 0$ the nondissipative state is also an equilibrium state [or stationary state, if $H=H(t)$]. When $[B(0),H]\ne 0$, instead, the unitary evolution in Eq.\ (\ref{Unitary}) can be regarded as a (constant entropy) limit cycle (or ``ridge'' as termed in Ref.\ \cite{Gheorghiu}), which coincides with the usual periodic solutions of the Schr\"odinger equation when $B$ is a one-dimensional projector, $B=|\psi\rangle\langle\psi|$ [$\Tr (B)=1$].

Indeed, Equation (\ref{EquationSingle}) preserves the rank and
nullity of $\rho$ \cite{Cimento} {or, as said in Ref.\ \cite{Gheorghiu},
preserves the cardinality of the set of nonzero eigenvalues of
$\rho$.
Because the equation of motion attracts the state operator towards
the highest entropy state or limit cycle compatible with the
initial mean values of the generators of the motion and the number
of zero eigenvalues, a minor perturbation of the state that
changes an initially zero eigenvalue to an arbitrarily small
nonzero value would cause an irreversible departure of the state
towards a different (higher entropy) equilibrium state or limit
cycle.  Hence, as long as there are zero eigenvalues of $\rho$,
i.e., unless $B=I$, all equilibrium states and limit cycles are
unstable.  Without this conclusion, we could not claim that the equation
of motion entails the second law of thermodynamics (see Appendixes
\ref{criteria} and \ref{stability}}, and the original papers, for
further discussion of this important point).

The Hamiltonian operator may be set to vary with time in order to model (adiabatic) energy exchange with the surroundings. Then the unitary part of the equation of motion, $-i[H(t),\rho]/\hbar$, may cause energy exchange, but not entanglement nor entropy exchange with the surroundings.  The dissipative term, instead, describes internal relaxation only and does not contribute to energy change in any case.

\section{Highest-entropy equilibrium states}\label{StableEquilibrium}

The only globally stable equilibrium states of the dynamics generated by Equation \ref{EquationSingle} are
\begin{equation}\label{stableeq}
\rho =\frac{ \exp(C)}{\Tr \exp(C)} =\frac{ \exp(-\beta H+\sum_i\nu_i G_i)}{\Tr \exp(-\beta H+\sum_i\nu_i G_i)} \ .
\end{equation}
The definition of global stability, which is stronger than
Lyapunov or local stability and is required by the second law
\cite{Lyapunov,Books}, is given and discussed in Appendix
\ref{stability}. The proof was discussed in Refs.\
\cite{Cimento,Attractor} and relied on a technical conjecture
discussed in Ref.\ \cite{Lyapunov} and later found to be proved by Hiai, Ohya, and Tsukada \cite{Hiai}.

For a system with generators of the motion $I$, $H$, $N_i$, Eq.\ (\ref{stableeq}) represents in general the grand canonical thermodynamic equilibrium states.  It reduces to the canonical equilibrium states if $N_i=n_iI$ for all $i$'s (to the microcanonical if also $H=eI$).

As is well known, states given by Eq.\ (\ref{stableeq}) are solutions of the constrained maximization problem
\begin{mathletters}\begin{eqnarray}
&&{\rm max}\ s(\rho)\\
&&{\rm subject\ to\ } r_i(\rho)=r_i \ {\rm and\ } \rho \ge 0 \ ,
\end{eqnarray}\end{mathletters}
where $s(\rho)=-\Boltz\Tr(\rho\ln\rho)$, $r_1(\rho)=\Tr(\rho I)$, $r_2(\rho)=\Tr(\rho H)$, $r_i(\rho)=\Tr(\rho G_i)$, $r_1=1$, $r_2$ and $r_i$'s given.

The inequality constraint $\rho\ge 0$ can be eliminated recasting the problem in terms of $\sq$,
\begin{mathletters}\label{maxproblem1}\begin{eqnarray}
&&{\rm max}\ s(\sq)\\
&&{\rm subject\ to\ } r_i(\sq)=r_i  \ ,
\end{eqnarray}\end{mathletters}
where $s(\sq)=-\Boltz\Tr [(\sq)^2\ln(\sq)^2]=s(\rho)$ and $r_i(\sq)=\Tr[(\sq)^2 R_i]=r_i(\rho)$.  The method of Lagrange multipliers then gives the condition (necessary but not sufficient)
\begin{equation}
\pardsq{s(\rho)}-\sum_i \lambda_i \pardsq{r_i(\rho)}=0 \ ,
\end{equation}
which, using Eqs.\ (\ref{gradR}) and (\ref{gradS}), becomes
\begin{equation}\label{necessary}
-2\Boltz\sq\ln\rho-2\Boltz\sq- \sum_i \lambda_i (\sq R_i+R_i\sq )=0 \ .
\end{equation}
It is noteworthy that Eq.\ (\ref{necessary}) is satisfied with obvious
identification of multipliers by Eq.\ (\ref{Cnondiss}),  where
$\sum_i\lambda_i\sq R_i$  commutes with $\sq\ln\rho$, i.e., with
$\rho$. Therefore, each nondissipative state satisfies the necessary
condition (\ref{necessary}) although it is not a solution of the
maximization problem (\ref{maxproblem1}) unless $B=I$.

\section{Onsager's reciprocity and \\ Callen's fluctuation-dissipation nonequilibrium relations}
\label{reciprocity}

 First, I introduce a general representation of
state operators particularly useful for representing nonequilibrium states. Any $\rho$ can be written as \cite{Onsager}
\begin{equation}\label{anystate}
\rho =\frac{ B\exp(-\sum_j f_j X_j)}{\Tr B\exp(-\sum_j f_j X_j)} \ ,
\end{equation}
where the hermitian operators in the set $\{I,X_j\}$ span the real
space $\Ell_h(\Hil)$ of linear hermitian operators on $\Hil$, and
$B$ is the usual idempotent operator defined in Eq.\ (\ref{Bdef}).
Indeed, the operator $B\ln\rho$ is always well-defined and belongs to
$\Ell_h(\Hil)$, so that
\begin{equation}\label{Blnrho}
B\ln\rho=-f_0 I-\sum_j f_j X_j= -f_0 BI-\sum_j f_j BX_j
\end{equation}
where the second equality follows from $B^2=B$. Therefore,
\begin{eqnarray}\label{sqlnrho}
&\sq\ln\rho=-f_0 \sq-\sum_j f_j \sq X_j  \ ,&\\
& x_j (\rho)=\Tr (\rho X_j)\ ,&\\
&s(\rho)=\Boltz f_0 + \Boltz\sum_j f_j\, x_j (\rho) \label{DDts} \ ,&\\
& D= - \sum_j f_j \left[\sq X_j- \sum_{k=1}^a (\sq X_j |\sq A_k) \sq A_k\right]\label{DAf}&
\end{eqnarray}
where, for simplicity, we use the expression (\ref{DA}) for $D$ in terms of the orthonomal set $\{\sq A_{i } \}$ [an equivalent expression in terms of Gram determinants is readily obtained using Eq.\ (\ref{sqlnrho}) in Eq.\ (\ref{Dgram}c)].

Notice from Eq.\ (\ref{DDts}) that the partial derivative
\begin{equation}
\Boltz f_j = \left.\frac{\partial s(\rho)}{\partial
x_j(\rho)}\right|_{x_{i\ne j}(\rho)}\
\end{equation}
may be interpreted as the generalized affinity or force representing the entropy change that corresponds to an independent change in the mean value of the linear observable $X_j$.

Next, we define the dissipative rate of change of the linear mean-value functional associated with operator $X_i$,
\begin{equation}\label{dissratexi}
\DDt{x_i(\rho)} =\Tr[\D_1(\rho,H,G_i) X_i]=2\!\left.\left(\sqdotD\right|\sq X_i\right)\; .
\end{equation}
Using $\sqdotD =-D/2\tau(\rho)$ [Eq.\ (\ref{rhodotD})] and Eq.\ (\ref{DAf}) yields the linear interrelations between dissipative rates and generalized affinities,
\begin{equation}\label{dissrate}
\DDt{x_i(\rho)} =\sum_j f_j \, L_{ij}(\rho)\ ,
\end{equation}
where the coefficients may be interpreted as generalized (Onsager) dissipative conductivities and are nonlinear functions of $\rho$,
\begin{mathletters}\label{Lij}\begin{eqnarray}
&L_{i j}(\rho) = L_{ji}(\rho)={\displaystyle \frac{1}{\tau(\rho)}}\!\left[(\sq X_i|\sq
X_j) \vphantom{\sum_{k=1}^a }\right.&\nonumber\\ & - \left.\sum_{k=1}^a (\sq X_i |\sq A_k)(\sq
A_k|\sq X_j)\right]& \\ &={\displaystyle \frac{1}{\tau(\rho)}}
\!\left[\cov{X_i}{X_j}- \sum_{k=1}^a
\cov{X_i}{A_k}\cov{A_k}{X_j}\right]& \\
&={\displaystyle \frac{1}{\tau(\rho)}} \!\left.\left([\sq X_i ]_{\bot\Ell {\{\sq
R_{i } \}}}\right|[\sq X_j ]_{\bot\Ell {\{\sq R_{i }
\}}}\right)\label{gramL} &\\
&={\displaystyle \frac{1}{\tau(\rho)}} \frac{\left|
\begin{array}{ccccc}
{\scriptstyle\cov{X_i}{X_j}} &{\scriptstyle\cov{R_1}{ X_j }} &\!\cdots\!& {\scriptstyle\cov{R_k}{ X_j }} & \!\cdots\!\\ \\
{\scriptstyle\cov{ X_i }{R_1}} &{\scriptstyle\cov{R_1}{R_1}} &\!\cdots\!& {\scriptstyle\cov{R_k}{R_1}} & \!\cdots\!\\
\vdots &\vdots & \!\ddots\! & \vdots&\!\ddots\!\\
{\scriptstyle\cov{ X_i }{R_k}} &{\scriptstyle\cov{R_1}{R_k}} &\!\cdots\!& {\scriptstyle\cov{R_k}{R_k}} & \!\cdots\!\\
\vdots &\vdots & \!\ddots\! & \vdots&\!\ddots\!
\end{array}
\right|
}{\displaystyle \Gamma(\{\sq R_k\})} \ .&
\end{eqnarray}\end{mathletters}
This equations are at the same time a proof and a generalization of Onsager's reciprocity relations \cite{Lars} extended to all nonequilibrium (dissipative) states.  Moreover, they give explicit expressions for the nonlinear dependence of the dissipative conductivities $ L_{ij}(\rho)$ on the state operator $\rho$, the generators of the motion and the internal-relaxation-time $\tau(\rho)$.

Using Eqs.\ (\ref{rateSfinal}) and (\ref{sqlnrho}), the rate of entropy change may be rewritten as a quadratic form of the generalized affinities,
\begin{equation}\label{sdotL}
\ddt{s(\rho)} =\Boltz\sum_i\sum_j f_i f_j L_{i j}(\rho)\ .
\end{equation}

The symmetric matrix $[ \{L_{i j}(\rho)\}]$ is a Gram matrix [see Eq.\ (\ref{gramL})] and, as such, it is nonegative definite, i.e., its determinant
\begin{equation}
\det[\{L_{ij}(\rho)\}] \ge 0 \ ,
\end{equation}
with strict positivity only if all operators
\begin{equation}
 [\sq X_i ]_{\bot\Ell {\{\sq R_{i } \}}}
\end{equation}
are linearly independent, in which case Eq.\ (\ref{dissrate}) may be solved to yield
\begin{equation}
f_j=\sum_i L^{-1}_{i j}(\rho)\DDt{x_i(\rho)}
\end{equation}
and the rate of entropy change can be written as a quadratic form of the dissipative rates
\begin{equation}
\ddt{s(\rho)} =\Boltz\sum_i\sum_j L^{-1}_{ij}(\rho)\DDt{x_i(\rho)} \DDt{x_j(\rho)}\ .
\end{equation}

Equations (\ref{Lij}) are also a proof and generalization of Callen's fluctuation-dissipation theorem \cite{Callen} extended to all nonequilibrium (dissipative) states.  Indeed, we interpret $ \cov{X_i}{X_j}$ as the codispersion (covariance) of measurement results of observables $X_i$ and $X_j$ when the system is in state $\rho$ and $ \cov{X_i}{X_i}$ as the dispersion (or fluctuations) of measurement results of observable $X_i$.  As in Callen's fluctuation-dissipation theorem, the expressions in Eq.\ (\ref{Lij}) relate codispersions with generalized conductivities.

A judicious choice of the set  $\{\sq X_{j } \}$ may greatly simplify these relations.  In particular, if $\{\sq X_{j } \}$ is any orthogonal extension of the orthonormal subset $\{\sq A_{i } \}$, in the sense that $(\sq X_j|\sq A_i)=0$ for all $A_i$'s and $j>a$, then for all $i$ and $j>a$,
\begin{equation}
L_{ij}(\rho) =\frac{1}{\tau(\rho)}\cov{X_i}{X_j}\ ,
\end{equation}
which relates directly [through the internal-relaxation time $\tau(\rho)$] covariances and fluctuations in the observables $X_i$ with their associated dissipative conductivities.

Onsager's \cite{Lars} and Callen's \cite{Callen} theorems are keystones of our understanding of irreversibility.  Indeed, the proof I proposed emerges not from the analysis of the Hamiltonian term of the equation of motion supplemented with the so-called assumption of microscopic reversibility, but from the dissipative term of the equation of motion, i.e., the only term responsible for irreversibility, with no additional assumptions.

Onsager's result \cite{Lars} was obtained from empirical observations on nonequilibrium
phenomena very close to stable thermodynamic equilibrium, so that the list of
$X_i$'s was indeed very short, and the result valid only for a limited class of
states.  Our result generalizes the validity of Onsager's reciprocity
relations to all nonequilibrium states, close and far from stable thermodynamic
equilibrium.  Of course, the price we have to pay to describe nonequilibrium
states far from stable equilibrium is that we must use a much larger,
possibly infinite list of $X_i$'s in Eq.\ (\ref{anystate}).

\section{Internal relaxation time vs time--energy Heisenberg uncertainty: maximal-entropy-production-rate ansatz}\label{Heisenberg}

By analogy with what done for the Hamiltonian characteristic time $\tau_H(\rho)$, we define the shortest characteristic time of the dissipative rate of change of $\sq$ by the relation (see Appendix \ref{times})
\begin{equation}\label{normD}
\frac{1}{\tau_D(\rho)^2}=4\left(\left.\sqdotD\right|\sqdotD\right) \,
\end{equation}
From Eqs.\ (\ref{sqdotD}) and (\ref{Dgram}) and $(\sq|\sq)=1$ [or,
directly, from Eq.\ (\ref{rateSfinal})], we find
\begin{equation}\label{tauD}
\tau_D(\rho)^2=\Boltz^2\tau(\rho)^2 \frac{\Gamma(\{\sq
R_i\})}{\Gamma(\sq S,\{\sq R_i\})}= \frac{\tau(\rho)^2}{(D|D)}\ .
\end{equation}

From this relation we may extract a possible ansatz on the functional form or at
least a lower bound for the internal-relaxation-time functional
$\tau(\rho)$ in Eq.\ (\ref{EquationSingle}) by
assuming that also $\tau_D(\rho)$, like $\tau_H(\rho)$, should satisfy
the Heisenberg uncertainty relation
\begin{equation}\label{normH}
\tau_D(\rho)^2\cov{H}{H}\ge\hbar^2/4 \ .
\end{equation}
This implies
\begin{equation}\label{taumin1}
\tau (\rho)^2 \ge \frac{\hbar^2}{4\Boltz^2}\frac{\Gamma(\sq S,\{\sq R_i\})}{\cov{H}{H}\,\Gamma(\{\sq R_i\})}= \frac{\hbar^2 (D|D)}{4\cov{H}{H}}
\end{equation}
or, using Eqs.\ (\ref{rateSfinal}) and (\ref{sdotL}),
\begin{equation}\label{taumin2}
\tau (\rho) \ge \frac{\hbar^2}{4\cov{H}{H}}\sum_i\sum_j f_i f_j L_{i j}(\rho)
\end{equation}
Relation (\ref{taumin1}) implies an upper bound to the rate of entropy production,
\begin{equation}\label{uppersdot}
\ddt{s(\rho)}\le \frac{2}{\hbar}\sqrt{\frac{\cov{H}{H}\,\Gamma(\sq S,\{\sq R_i\})}{\Gamma(\{\sq R_i\})}} \ .
\end{equation}
By analogy with the Hamiltonian time, a possible ansatz that I propose should be carefully considered is that also $\tau_D(\rho)$ and, therefore, $\tau (\rho)$ be equal exactly to the lower bound [i.e., strict equality in Eqs.\ (\ref{taumin1}), (\ref{taumin2}) and (\ref{uppersdot})], corresponding to the maximal entropy production rate compatible with the time--energy uncertainty relation.  This corresponds to a truly maximal entropy production dynamics. The square-root state operator $\sq$ not only moves in the direction of steepest entropy ascent, but it does so at the highest rate compatible with the uncertainty principle.

This ansatz is very intriguing, conceptually appealing and fraught with far-reaching physical consequences. But, like any other hypotheses on the functional form of $\tau(\rho)$, it should be verified against known behavior in worked-out specific cases, such as the examples considered in Refs.\ \cite{Fluorescence,Korsch,Gheorghiu} and the variety of physical problems for which nonlinear modifications of the Schr\"odinger equation have been deemed necessary but have not yet been resolved \cite{nonlinearS}.

If proved valid, this ansatz would complete the dynamics without need of additional physical constants and would provide a fundamental microscopic foundation of the macroscopic observation that Nature always evolves at the fastest possible rate along the most direct path towards maximum entropy compatible with the system's structure and external constraints.  Interpreted in this way, my nonlinear dynamics would provide a unifying fundamental microscopic foundation of all phenomenological theories of irreversible processes advanced in the last seventy years after the pioneering work of Onsager.

\section{Variational principle formulation}\label{variational}

Following Gheorghiu-Svirschevski \cite{Gheorghiu},  the direction of steepest-entropy-ascent can also be found by considering the constrained maximization problem
\begin{mathletters}\label{maxprob}\begin{eqnarray}
&&{\rm max}\ \ddt{s}(\sqdot)\\
&&{\rm subject\ to\ } \ddt{r_i}(\sqdot)=\dot r_i \ {\rm and\ }  (\sqdot | \sqdot)=\frac{1}{\tau_D^2}\ ,
\end{eqnarray}\end{mathletters}
where $\dot r_i$ and $\tau_D^2$ are given real functionals of $\rho$.  The necessary solving condition in terms of Lagrange multipliers is
\begin{equation}
\pardsqdot{}\ddt{s}-\sum_i \lambda_i \pardsqdot{}\ddt{r_i}-\lambda_0 \pardsqdot{} (\sqdot | \sqdot) =0 \ .
\end{equation}
Using Eqs.\ (\ref{rateR}) and (\ref{rateS}), it becomes
\begin{equation}\label{maxS}
\pardsq{s(\rho)}-2\sum_i \lambda_i \sq R_i-\lambda_0 \sqdot=0 \ ,
\end{equation}
where the multipliers must be determined by substitution in the system of constraint equations.

It easy to verify that our expression of $\sqdotD=-D/2\tau(\rho)$ in Eqs.\ (\ref{sqdotD}) and (\ref{Dgram}) is an explicit solution of Eq.\ (\ref{maxS})  in the case $\dot r_i=0$ and $\tau_D$ given by the explicit expression in Eq.\ (\ref{tauD}).  See Appendix \ref{ExtensionRates} for $\dot r_i\ne 0$.

\section{BLOCH NEAR-EQUILIBRIUM LINEAR LIMIT}\label{Bloch}

In Ref.\ \cite{Cimento} we have shown that in the neighbourhood of each equilibrium state $\rho_{\rm e}$ \{given by Eq.\ (\ref{Nondissipative}) with the additional condition $[B,H]=0$\}, if we assume $\tau(\rho)$ constant (e.g., $\tau_{\rm e}=\tau(\rho_{\rm e})\ne 0$), the state operators in the subset with $B=B_{\rm e}$ and $\Tr(\rho R_i)= \Tr(\rho_{\rm e} R_i)$ (i.e., with the same nullity, rank and mean values of the generators as the equilibrium state) obey a linearized form of the equation of motion that has the form of a Bloch relaxation equation,
\begin{equation}\label{blocheq}
\ddt{\rho}\to -\frac{i}{\hbar}[H,\rho]-\frac{\rho-\rho_{\rm e}}{\tau_{\rm e}} \ ,
\end{equation}
so that the solution of the equation of motion is
\begin{eqnarray}\label{relaxation}
\rho(t)&\to& \exp(-t/\tau_{\rm e}) U(t)\rho(0) U^{-1}(t)\nonumber\\&+& \left[1-\exp(-t/\tau_{\rm e}) \right] \rho_{\rm e} (0) \ ,
\end{eqnarray}
with $U(t)$ given by Eq.\ (\ref{Bunitary}) \{note that if there are no non-Hamiltonian generators the condition $[B,C]=0$ implies $[\rho,H]=0$ and, therefore, $U(t)\rho(0) U^{-1}(t)= \rho(0)$\}.

Gheorghiu-Svirschevski \cite{Gheorghiu}, derived a linearized form for the case with no non-Hamiltonian generators valid also when $[\rho,H]\ne 0$ (but assuming $\tau\to\tau_{\rm e}\ne 0$), and showed that   it yields a generalized Fokker-Planck equation for a free particle, and a Langevin equation for a harmonic oscillator.

It is however noteworthy that such linearized limit behavior should be revised if we take $\tau(\rho)$ equal to the lower bound imposed by the time--energy uncertainty relation [Eq.\ (\ref{taumin1}) with strict equality], because then, if $\cov{H}{H}\ne 0$, $\tau(\rho)\to 0$ as $\rho\to\rho_{\rm e}$.

\section{DIVISIBLE SYSTEMS DYNAMICS.  DEFINITIONS AND NOTATION}\label{Composite}

The composition of
the system is embedded in the structure of the Hilbert space as a direct product of the subspaces associated with the individual elementary constituent subsystems, as well as in the form of the Hamiltonian operator.

In this section, we consider a system composed of $M$ distinguishable and indivisible elementary constituent subsystems. For example, each subsystem may be a different elementary particle or a Fermi-Dirac or Bose-Einstein field (in which case the corresponding $\Hil^\J$ is a Fock space).  The Hilbert space is
\begin{equation}
\label{HilbertSpace}\Hil = \Hil^1\Otimes \Hil^2\Otimes\cdots\Otimes\Hil^M \ ,
\end{equation}
and the Hamiltonian operator
\begin{equation}
\label{Hamiltonian} H  = \sum_{J=1}^M H_\J\Otimes I_\Jbar + V \ , \end{equation}
where $H_\J$ is the Hamiltonian operator on $\Hil^\J$ associated with the $J$-th subsystem when isolated and $V$ (on $\Hil$) the interaction Hamiltonian among the $M$ subsystems.

For convenience, we denote by $\Hil^\Jbar$ the direct product of the Hilbert spaces of all subsystems except the $J$-th one, so that the Hilbert space of the overall system
\begin{equation}\Hil=\Hil^\J\Otimes\Hil^\Jbar\end{equation}
and the identity operator $I = I_\J\Otimes I_\Jbar$.

The
subdivision into elementary constituents, considered as indivisible, and reflected by
the structure of the Hilbert space $\Hil$ as a direct product of subspaces, is particularly
important because it defines the level of description of the system and
specifies its elementary structure.  This determines also the structure of
the nonlinear dynamical law I proposed. In a sense, this is a price we have to pay in order to free ourselves from the assumption of linearity of the law of motion.

In other words, the form of the superoperators $\D_M$ and $\D_1$ are different depending on whether the system is or is not subdivisible into indivisible subsystems.  For example, consistently with the conditions listed in Appendix \ref{criteria}, we request that the superoperator $\D_M$ satisfies the strong separability conditions \cite{Korsch,Czachor}
\begin{mathletters}\label{separability}\begin{eqnarray}
&&\D_M(\rho_A\Otimes\rho_B,H_A\Otimes I_B\!+\!I_A\Otimes H_B,
G_{iA}\Otimes I_B\!+\!I_A\Otimes G_{iB})\nonumber\\ &&
=\D_{M_A}\!(\rho_A, \! H_A, \! G_{iA})\Otimes \rho_B \!+
\!\rho_A\Otimes\D_{M_B}\!(\rho_B,\! H_B,\!
G_{iB})\label{weakseparability},\\ &&\Tr_B [\D_M(\rho,H_A\Otimes
I_B\!+\!I_A\Otimes H_B, G_{iA}\Otimes I_B\!+\!I_A\Otimes
G_{iB})]\nonumber\\ &&\qquad =\F_{M_A}\!(\rho, \! H_A, \! G_{iA})
\qquad {\rm for\ any}\ \rho\label{strongseparabilityA},\\ & &\Tr_A
[\D_M(\rho,H_A\Otimes I_B\!+\!I_A\Otimes H_B, G_{iA}\Otimes
I_B\!+\!I_A\Otimes G_{iB})]\nonumber\\ && \qquad=\F_{M_B}\!(\rho,
\! H_B, \! G_{iB}) \qquad {\rm for\ any}\
\rho\label{strongseparabilityB},
\end{eqnarray}\end{mathletters}
where subsystems $A$ and $B$ are obtained by partitioning the set
of constituents 1, 2, \dots, $M$ into two disjoint subsets of
$M_A$ and $M_B$ constituents, respectively ($M_A+M_B=M$). Note
that, of course, if $\rho_A\Otimes\rho_B$ then $\F_{M_A}\!(\rho,
\! H_A, \! G_{iA})$ $=\D_{M_A}\!(\rho_A, \! H_A, \! G_{iA})$ and
$\F_{M_B}\!(\rho, \! H_B, \! G_{iB})$ $=\D_{M_B}\!(\rho_B, \! H_B,
\! G_{iB})$.

It is noteworthy that trying to apply the same conditions to
superoperator $\D_1$ would be physically meaningless, because if
the system were subdivisible then $\D_1$ would not be its
dissipative evolution superoperator.

 Conditions
(\ref{strongseparabilityA}) and (\ref{strongseparabilityB})
prevent locality problems because they guarantee that changes of
the Hamiltonian (or the other generators of the motion) in one of
two noninteracting subsystems cannot affect the mean values of
local observables of the other subsystem.  For example, assume
that subsystems $A$ and $B$ are correlated but not interacting. We
may switch on a measurement apparatus within $B$ and therefore
alter the Hamiltonian $H_B$.  By virtue of Eqs.
(\ref{strongseparabilityA}) and (\ref{strongseparabilityB}), the
rate of change of the reduced state operator $\rho_A=\Tr_B(\rho)$
does not depend on $H_B$ and, therefore, all functionals of
$\rho_A$ (local observables) remain unaffected by the change in
$B$, i.e., no faster-than-light communication can occur between
$B$ and $A$ (of course, if we exclude the projection postulate).

In addition, we must consider the following additional non-trivial conditions of separate energy conservation of noninteracting subsystems for any $\rho$ (see Appendix \ref{criteria}, Condition 6)
\begin{mathletters}\label{separabilityenergy}\begin{eqnarray}
\Tr[(H_A\Otimes I_B )\,\D_M(\rho,H_A\Otimes I_B\!+\!I_A\Otimes H_B, G_i)]&=& 0 \ ,\\
\Tr[(I_A\Otimes H_B )\,\D_M(\rho,H_A\Otimes I_B\!+\!I_A\Otimes H_B, G_i)]&=& 0 \ ,
\end{eqnarray}\end{mathletters}
and of separate entropy nondecrease for uncorrelated (possibly interacting) subsystems (see Appendix \ref{criteria}, Condition 7)
\begin{mathletters}\label{separabilityentropy}\begin{eqnarray}
\Tr[(S_A\Otimes I_B )\,\D_M(\rho_A\Otimes\rho_B,H, G_i)]&\ge& 0 \ ,\\
\Tr[(I_A\Otimes S_B )\,\D_M(\rho_A\Otimes\rho_B,H, G_i)]&\ge& 0 \ ,
\end{eqnarray}\end{mathletters}
where $S_A=-\Boltz B_A\ln\rho_A$, $S_B=-\Boltz B_B\ln\rho_B$.

For each type of particle in the system, we may write without loss of generality the number-of-particles-of-$i$-th-type operator associated with the system as \cite{Note}
\begin{equation}
\label{Number} N_i = \sum_{J=1}^M N_{i\J}\Otimes I_\Jbar \qquad {\rm for}\  i = 1,2,\dots,r \ ,\end{equation}
where  $N_{i\J}$ denotes
the number-of-particles-of-$i$-th-type operator associated with the $J$-th subsystem.

In general, we assume that the set of linear hermitian operators $I$, $H$, $G_i$ on $\Hil$, always including $I$ and $H$, are the generators of the motion of the composite system.  For example, the list of non-Hamiltonian generators $G_i$ may coincide with that of the number operators defined in Eq.\ (\ref{Number}).

For each subsystem $J$ we denote by $\Ell (\Hil^\J)$ the space  of linear operators on $\Hil^\J$ equipped with the real scalar product
\begin{equation}
\label{ScalarJ} (F_\J|G_\J)_\J = \frac{1}{2}\Tr_\J \left(F_J ^\dagger G_\J + G_J ^\dagger F_\J\right)  \ .
\end{equation}

For a given state operator $\rho$ on $\Hil$, given linear hermitian $F$ and $G$ on $\Hil$, and each subsystem $J$, in addition to that already defined in Eqs.\ (\ref{Scalar}), (\ref{Delta}) and (\ref{Cov}), we define  the following convenient notation \cite{Cimento,Onsager}
\begin{eqnarray}
\label{reducedstate} \rho_\J &=&\Tr_\Jbar(\rho) \ , \\
\label{reducedotherstate} \rho_\Jbar &=& \Tr_\J (\rho) \ , \\
\label{privateF}(F)^\J&=&\Tr_\Jbar [(I_\J\Otimes \rho_\Jbar) F]\ , \\
\label{CrossJsup} \cov{F}{G}^\J &=& (\sqJ (\Delta F)^\J | \sqJ (\Delta G)^\J)_\J\nonumber \\&=& \frac{1}{2}\Tr_\J (\rho_\J \{(\Delta F)^\J, (\Delta G)^\J \}) \ ,
 \end{eqnarray}
where $\Tr_\Jbar$ denotes the partial trace over $\Hil^\Jbar$, $\Tr_\J$ over $\Hil^\J$, $\rho_\J$ is the reduced state operator of elementary subsystem $J$ and $\rho_\Jbar$ that of the composite of all other subsystems.

In view of the special role they play in the equation of motion, we interpret the operators $(H)^\J$ and $(S)^\J$,
\begin{eqnarray}
\label{privateH}&(H)^\J=\Tr_\Jbar [(I_\J\Otimes \rho_\Jbar) H]\ ,& \\
\label{privateS}&(S)^\J=\Tr_\Jbar [(I_\J\Otimes \rho_\Jbar) S]\ , &
\end{eqnarray}
respectively, as ``internal perception'' operators representing the way the overall-system Hamiltonian and entropy operators are ``felt'' locally within the $J$-th constituent subsystem.

For a given $\rho_\J$, and given linear hermitian $F_\J$ and $G_\J$ on $ \Hil^\J$, we further define the notation
\begin{eqnarray}
\label{CrossJsub} \cov{F_\J}{G_\J}_\J &=& (\sqJ \Delta F_\J | \sqJ \Delta G_\J)_\J\nonumber \\&=&  \frac{1}{2}\Tr_\J(\rho_\J \{\Delta F_\J, \Delta G_\J \}) \ , \\
\label{sqrtJ} &&\sqJ \ , \\
\label{BJ} &&B_\J \ , \\
\label{SJ} S_\J&=&- \Boltz B_\J \ln \rho_\J \ ,
\end{eqnarray}
where the operators $\sqJ$, $B_\J$ and $S_\J$ are obtained by substituting each nonzero eigenvalue $p_i$ of $\rho_\J$ with its
positive square root, unity  and $-\Boltz \ln p_i$.  In general, $S_\J\ne(S)^\J$.

The entropy is defined for a subsystem $J$ only if it is not correlated with the other subsystems, i.e., if $\rho$ can be written as
\begin{equation}
\rho = \rho_\J\Otimes \rho_\Jbar \ ;
\end{equation}
then the subsystem entropy is given by the nonlinear state functional of the reduced state operator,
\begin{eqnarray}
\label{Subentropyfunct} &s_\J(\rho_\J) = \Tr_\J (\rho_\J S_\J)=
-\Boltz \Tr_\J (\rho_\J\ln\rho_\J) &\nonumber\\&= \Tr[\rho (S_\J\Otimes I_\Jbar)]=   (\sqJ |\sqJ
S_\J)_\J \ ,&\\ \label{Subentropyop} &S = S_\J\Otimes I_\Jbar +
I_\J\Otimes S_\Jbar \ ,&\\ \label{entropyadditivity} &s(\rho) =
s_\J(\rho_\J) + s_\Jbar(\rho_\Jbar) \ ,&
\end{eqnarray}
and we also have $ (\Delta S)^\J = \Delta S_\J $.

When subsystem $J$ is correlated, instead, its entropy is not defined; however, the functional
\begin{equation}
\label{Subentropyperception} s^\J(\rho) = \Tr_\J [\rho_\J(S)^\J]=
(\sqJ |\sqJ (S)^\J)_\J
\end{equation}
may be interpreted as the subsystem's local ``internal perception''
of the overall-system entropy.
Only when $J$ is uncorrelated, $ s^\J(\rho)=s(\rho)= s_\J(\rho_\J) + s_\Jbar(\rho_\Jbar)$.

The energy is defined for a subsystem $J$ only if it is not interacting with the other subsystems, i.e., if $H$ can be written as
\begin{equation}
H = H_\J\Otimes I_\Jbar + I_\J \Otimes H_\Jbar \ ;
\end{equation}
then it is given by the mean-value state functional
\begin{equation}
\label{Subenergy} e_\J(\rho_\J) =\Tr_\J (\rho_\J H_\J)=(\sqJ |\sqJ H_\J)_\J \ ,
\end{equation}
and we also have
$(\Delta H)^\J = \Delta H_\J$ \ .

When subsystem $J$ is interacting, instead, its energy is not defined, however the functional
\begin{equation}
\label{Subenergyperception} e^\J(\rho) = \Tr_\J [\rho_\J(H)^\J]= (\sqJ |\sqJ (H)^\J)_\J
\end{equation}
may be interpreted as the subsystem's local ``internal perception''
of the overall-system energy.
Only when $J$ is non-interacting, $ e^\J(\rho)=e(\rho)= e_\J(\rho_\J) + e_\Jbar(\rho_\Jbar)$.

The number-of-particles-of-$i$-th-type of the overall system in state $\rho$ and of each subsystem $J$ are given by the mean-value state functionals
$n_i(\rho) =\Tr (\rho N_i)$
and
$ n_{i\J}(\rho)= \Tr [\rho (N_{i\J}\Otimes I_\Jbar)]= n_{i\J}(\rho_\J)= \Tr_\J (\rho_\J N_{i\J})= (\sqJ |\sqJ N_{i\J})_\J $, respectively, and from Eq.\ (\ref{Number}) we clearly have $ n_i(\rho)=\sum_{J=1}^M n_{i\J}(\rho_\J)$.
It is noteworthy that, for any $\rho$,
\begin{equation}
\label{NiJ} (N_{i\J}\Otimes I_\Jbar)^\J= N_{i\J} \ .
\end{equation}

For generality, we assume that the generators of the motion on $\Hil$ are $I$, $H$ and $G_i$, with $[G_i,H]=0$ but not necessarily with the separated structure of the number operators.  Moreover,
for each state $\rho$ and each constituent $J$, we denote by $\{R_{i\J}$, $i=0$, 1, 2, \dots, $z^\J$\} a set of hermitian operators in $\Ell(\Hil)$ such that the operators in the set \{$\sqJ (R_{i\J})^J$\}, where
\begin{equation}
\label{privateR}( R_{i\J})^\J=\Tr_\Jbar [(I_\J\Otimes \rho_\Jbar) R_{i\J}]\ ,
\end{equation}
are linearly independent and span the linear manifold generated by the operators $\sqJ I_\J$, $\sqJ (H)^\J$, $\sqJ (G_i)^\J$.  If the latter operators are linearly independent, for subsystem $J$, then the set $\{ R_{i\J}\}$ may be chosen to coincide with the generators of the motion.  If they are not independent, then it could be a smaller subset (in which it is convenient, though not necessary, to maintain $R_{0\J}=I $ and, if possible, $R_{1\J}=H $). In any case, we call the operators in the set \{$ R_{i\J}$\} the (instantaneous) ``generators of the motion of subsystem $J$''. The structure of the dissipative term $\D_M$ is invariant under transformation from one set \{$ R_{i\J}$\} to any other \{$ R'_{i\J}$\} with the same defining properties.

For each instantaneous generator $ R_{i\J}$ of subsystem $J$, we define the local ``internal perception'' mean functional
\begin{equation}
\label{Subgeneratorperception} r_{i\J}^\J(\rho) = \Tr_\J [\rho_\J(R_{i\J})^\J]= (\sqJ |\sqJ (R_{i\J})^\J)_\J \ .
\end{equation}

We denote by
\begin{equation}
\label{linearSpanM}\Ell {\{\sqrt{\rho_\J}(R_{i\J})^J\}}=
\Ell {\{\sqrt{\rho_\J}I_\J, \sqrt{\rho_\J} (H)^\J, \sqrt{\rho_\J} (G_i)^\J\}}
\end{equation}
the linear span of the operators \{$\sqJ (R_{i\J})^J$\} or, that is the same, the linear span of the operators $\sqJ I_\J$, $\sqJ (H)^\J$, $\sqJ (G_i)^\J $.

By definition of the set \{$ R_{i\J}$\}, the Gram determinant
\begin{eqnarray}
\Gamma (\{\sqJ (R_{i\J})^J\}) &=& \det [\{(\sqJ (R_{i\J})^\J|\sqJ R_{j\J})^\J)_\J \}] \nonumber\\
&=& \det [\{\cov{R_{i\J}}{ R_{j\J}}^\J \}]
\end{eqnarray}
is always strictly positive.

The set  $\{ R_{i\J} \}$  can be conveniently chosen so that $\{\sqJ (R_{i\J})^\J\}$ is (instantaneously) an orthonormal set, in which case we denote it by $\{\sqJ (A_{i\J})^\J\}$ and
\begin{eqnarray}
&(\sqJ (A_{i\J})^\J|\sqJ A_{j\J})^\J)_\J =\delta_{ij} \ , &\\
&\Gamma_\J (\{\sqJ (A_{i\J})^J\})=1 \ .&
\end{eqnarray}

\section{The steepest-entropy-ascent ansatz for a composite system}

Maintaining the validity of Eqs.\ (\ref{rhodot}) and (\ref{sqdot}), we now assume
\begin{equation}
\label{rhodotM}\sq\, \sqdotD =\sum_{J=1}^M \sqJ \,\sqdotJD \Otimes \rho_\Jbar\ ,
\end{equation}
as a first step to guarantee the strong separability condition (\ref{separability}) [see Appendix \ref{wrong} for a discussion related to the form of Eq.\ (\ref{rhodotM})].

The second step is to make sure that $\sqdotJD$, which in general may be a function of the operators $\rho$, $H$ and $G_i$ on $\Hil$, reduces to a function of the operators $\rho_\J$, $H_\J$ and $G_{i\J}$ on $\Hil^\J$ only, whenever the constituent is, at the same time, uncorrelated ($\rho=\rho_\J\Otimes \rho_\Jbar$), non-interacting ($H_\J \Otimes I_\Jbar \!+\!I_\J \Otimes H_\Jbar $) and not coupled through the non-Hamiltonian generators ($G_{i\J }\Otimes I_\Jbar \!+\!I_\J \Otimes G_{i\Jbar }$).

To preserve the formal analogy with the notation for the indivisible system, we define the local ``partial gradient'' operators
\begin{eqnarray}
\label{gradRJ}
\left[\pardsqJ{r_{i\J}^\J(\rho)}\right]_{(R_{i\J})^\J}&=& \sqJ
(R_{i\J})^\J+ (R_{i\J})^\J \sqJ \ ,\\ \label{gradSJ}
\left[\pardsqJ{s^\J(\rho)}\right]_{(S)^\J}&=& 2\sqJ (S)^\J \ .
\end{eqnarray}

Now, it is easy to show that the structure of ${\rm d}\rho/{\rm d}t$ assumed with Eqs. (\ref{rhodot}), (\ref{sqdot}) and (\ref{rhodotM}) yields \cite{NoteonBdot}
\begin{eqnarray}
\label{ratesRJ} \ddt{r_{i\J}(\rho)}&=&\Tr(
\ddt{\rho}R_{i\J})=2\sum_{J=1}^M \left.\left(\sqdotJD \right|\sqJ
(R_{i\J})^\J \right)_\J\ , \\ \label{ratesSJ}
\ddt{s(\rho)}&=&-\Boltz\Tr( \ddt{\rho})-\Boltz\Tr( \ddt{\rho}B\ln\rho)
\nonumber \\&=& \sum_{J=1}^M \left[-2\Boltz \left.\left(\sqdotJD
\right|\sqJ I_\J \right)_\J \right.\nonumber \\& &\qquad\left.+2
\left.\left(\sqdotJD \right|\sqJ (S)^ \J \right)_\J\right] \nonumber\\
&=& \sum_{J=1}^M \left[\vphantom{\left(\left.
\left[\pardsqJ{s^\J(\rho)}\right]_{(S)^\J}\right|\sqdotJD \right)_\J
}-2\Boltz \left(\sqJ I_\J \left|\sqdotJD \right.\right)_\J
\right.\nonumber \\& &\qquad\left. + \left(\left.
\left[\pardsqJ{s^\J(\rho)}\right]_{(S)^\J}\right|\sqdotJD
\right)_\J\right] \ .
\end{eqnarray}

Finally, we assume that each $\sqdotJD$ is: (A) orthogonal to $\Ell {\{\sqrt{\rho_\J}(R_{i\J})^J\}}$, so that all the rates of change in Eq.\ (\ref{ratesRJ}) and the first term in the rate of entropy change in Eq.\ (\ref{ratesSJ}) are zero; and (B) in the direction of the local partial gradient of the ``internally perceived'' overall-system entropy functional, $s^\J(\rho)$, i.e., we take
\begin{mathletters}\label{sqdotJD}\begin{eqnarray}
\sqdotJD&=&\frac{1}{4\Boltz\tau_\J(\rho)} \left[\left[\pardsqJ{s^\J(\rho)}\right]_{(S)^\J}\right]_{\bot \Ell {\{\sqrt{\rho_\J}(R_{i\J})^J\}}}\!\!\!\!\!\\
&=&-\frac{1}{2\tau_\J(\rho)} D_\J \ ,
\end{eqnarray}\end{mathletters}
where, similarly to what done for the indivisible system,
\begin{mathletters}\label{DJ}\begin{eqnarray}
&&D_\J=\left[\sqJ(B\ln\rho)^\J \right]_{ \bot\Ell {\{\sqrt{\rho_\J}(R_{i\J})^J\}}}\\
&&= \sqJ(B\ln\rho)^\J-\left[\sqJ(B\ln\rho)^\J \right]_{ \Ell {\{\sqrt{\rho_\J}(R_{i\J})^J\}}}\\ \nonumber & \ &\\
&&=-\frac{1}{\Boltz}\frac{\left|
\begin{array}{ccccc}
\sqJ (\Delta S)^\J & \sqJ (\Delta R_{1\J})^\J &\!\cdots\!& \sqJ (\Delta R_{i\J})^\J &\!\cdots\!\\ \\
{\scriptstyle\cov{S}{R_{1\J}}^\J }& {\scriptstyle\cov{R_{1\J}}{R_{1\J}}^\J } &\!\cdots\!& {\scriptstyle\cov{R_{i\J}}{R_{1\J}}^\J } & \!\cdots\!\\
\vdots & \vdots & \!\ddots\! & \vdots&\!\ddots\!\\
{\scriptstyle\cov{S}{R_{i\J}}^\J }& {\scriptstyle\cov{R_{1\J}}{R_{i\J}}^\J } &\!\cdots\!& {\scriptstyle\cov{R_{i\J}}{R_{i\J}}^\J } & \!\cdots\!\\
\vdots& \vdots & \!\ddots\! & \vdots&\!\ddots\!
\end{array}
\right|}
{\Gamma(\{\sqJ (R_{i\J})^\J \}) } \ .\label{DJc}
\nonumber\\
&\ &
\end{eqnarray}\end{mathletters}

All the results found for the single constituent extend to the composite system in a straightforward way.

The rate of entropy change
\begin{mathletters}\label{rateSfinalJ}\begin{eqnarray}
\ddt{s(\rho)}&=& \sum_{J=1}^M \frac{\Boltz }{ \tau_\J(\rho)} (D_\J|D_\J)_\J \\
&=&\sum_{J=1}^M {4\Boltz\tau_\J(\rho)}\left(\sqdotJD \left|\sqdotJD \right.\right)_\J \\
&=& \sum_{J=1}^M \frac{1 }{ \Boltz\tau_\J(\rho)} \frac{\Gamma(\sqJ (S)^\J,\{\sqJ (R_{i\J})^\J \})}{\Gamma(\{\sqJ (R_{i\J})^\J \})} \ .
\end{eqnarray}\end{mathletters}

The dynamics reduces to the Schr\"odinger-von Neumann unitary Hamiltonian dynamics when
\begin{mathletters}\label{CnondissJ}\begin{eqnarray}
&\sqJ(B\ln\rho)^\J = \sqJ ( C )^\J \ ,&\\
 &C=\sum_i \lambda_{i\J}R_{i\J}\ ,&
\end{eqnarray}\end{mathletters}
for each $J$ and, therefore, the state is nondissipative \{equilibrium if $[B,H]=0$, limit cycle if $[B,H]\ne 0$, including the case of pure-state standard QM when $\Tr (B)=1$\}.

The maximum entropy thermodynamic equilibrium states are given by Eq.\ (\ref{stableeq}).

Onsager's reciprocity relations follow again from Eq.\ (\ref{Blnrho}) for the operator $B\ln\rho$.  Equations (\ref{dissrate}) and (\ref{sdotL}) are still valid [of course, with $\D_M$ in Eq.\ (\ref{dissratexi})], with the dissipative conductivities given by
\begin{mathletters}\label{LijJ}\begin{eqnarray}
&&L_{i j}(\rho) = L_{ji}(\rho) ={\displaystyle \sum_{J=1}^M}
L^\J_{ij}(\rho)\ ,\\ &&L^\J_{ij}(\rho)=L^\J_{ji}(\rho)\nonumber\\
&&= \frac{1}{\tau_\J(\rho)} \frac{\left|
\begin{array}{ccccc}
{\scriptstyle\cov{X_i}{X_j}^\J } &{\scriptstyle\cov{R_1}{ X_j }^\J } &\!\cdots\!& {\scriptstyle\cov{R_k}{ X_j }^\J } & \!\cdots\!\\ \\
{\scriptstyle\cov{ X_i }{R_1}^\J } &{\scriptstyle\cov{R_1}{R_1}^\J } &\!\cdots\!& {\scriptstyle\cov{R_k}{R_1}^\J } & \!\cdots\!\\
\vdots &\vdots & \!\ddots\! & \vdots&\!\ddots\!\\
{\scriptstyle\cov{ X_i }{R_k}^\J } &{\scriptstyle\cov{R_1}{R_k}^\J } &\!\cdots\!& {\scriptstyle\cov{R_k}{R_k}^\J } & \!\cdots\!\\
\vdots &\vdots & \!\ddots\! & \vdots&\!\ddots\!
\end{array}
\right|
}{\displaystyle \Gamma(\{\sqJ (R_{i\J})^\J \})} \ .\nonumber \\
\end{eqnarray}\end{mathletters}

Callen's fluctuation-dissipation relations, implied by the explicit structure [Eq.\ (\ref{LijJ})] of the Onsager dissipative conductivities, are greatly simplified if we choose the orthonormal set $\{\sqrt{\rho_\J}(A_{i\J})^J\}$ instead of $\{\sqrt{\rho_\J}(R_{i\J})^J\}$ and the set $\{\sqrt{\rho_\J}(X_{j\J})^J\}$ to be an orthogonal extension of $\{\sqrt{\rho_\J}(A_{i\J})^J\}$; then,
\begin{equation}
L^\J_{ij}(\rho) = \frac{1}{\tau_\J(\rho)}\cov{X_i}{X_j}^\J\ .
\end{equation}

The conjecture about the time--energy Heisenberg  uncertainty
relation in Section \ref{Heisenberg} can be extended as well by
assuming that each $\sqdotJD$ gives rise to a characteristic time
$\tau_{D\J}(\rho)$ defined by (see Appendix \ref{times})
\begin{equation}\label{normDJ}
\frac{1}{\tau_{D\J}(\rho)^2}=4\left(\left.\sqdotJD\right|\sqdotJD\right)
\ ,
\end{equation}
which should independently satisfy the uncertainty relation
\begin{equation}\label{normHJ}
\tau_{D\J}(\rho)^2\cov{H}{H}^\J\ge\hbar^2/4 \ .
\end{equation}
Therefore, using Eqs.\ (\ref{sqdotJD}) and (\ref{DJ}), we obtain the lower bounds to the internal relaxation times,
\begin{equation}\label{taumin1J}
\tau_\J (\rho)^2\! \ge\! \frac{\hbar^2}{4\Boltz^2} \frac{\Gamma(\sqJ (S)^\J,\{\sqJ (R_{i\J})^\J \})}{\cov{H}{H}^\J\,\Gamma(\{\sqJ (R_{i\J})^\J \})}\!=\!\frac{\hbar^2 (D_\J|D_\J)_\J}{4\cov{H}{H}^\J }
\end{equation}
or, using, Eqs.\ (\ref{rateSfinalJ}) and (\ref{sdotL}),
\begin{equation}\label{taumin2J}
\tau_\J (\rho) \ge \frac{\hbar^2}{4\cov{H}{H}^\J}\sum_i\sum_j f_i f_j L^\J_{i j}(\rho) \ ,
\end{equation}
and the upper bound to the entropy production rate
\begin{equation}\label{uppersdotJ}
\ddt{s(\rho)}\!\le \!\sum_{J=1}^M \!\frac{2}{\hbar}\sqrt{\frac{\cov{H}{H}^\J\,\Gamma(\sqJ (S)^\J,\{\sqJ (R_{i\J})^\J \})}{\Gamma(\{\sqJ (R_{i\J})^\J \})} }\, .
\end{equation}

Taking strict equality in each Eq.\ (\ref{taumin1J}) yields the maximum entropy production rate compatible with the uncertainty relations.

The variational formulation can be found, after  assuming the
structure of Eq.\ (\ref{rhodotM}), by solving the constrained
maximization problem
\begin{mathletters}\label{maxprobJ}\begin{eqnarray}
&&{\rm max}\ \ddt{s}(\{\sqdotJ\})\\ &&{\rm subject\ to\ }
\ddt{r_{i\J}}(\{\sqdotJ\})=\dot r_{i\J} \ {\rm and\ }  (\sqdotJ |
\sqdotJ)=\frac{1}{\tau_{D\J}^2}\ ,\nonumber \\
\end{eqnarray}\end{mathletters}
where $\dot r_{i\J}$ and $\tau_{D\J}^2$ are given real functionals
of $\rho$. By virtue of Eqs.\ (\ref{ratesRJ}) and (\ref{ratesSJ}),
the necessary solving conditions in terms of Lagrange multipliers
for each $\sqdotJ$ become
\begin{equation}\label{maxSJ}
\left[\pardsqJ{s^\J(\rho)}\right]_{(S)^\J}\!\!-2\sum_i \lambda^\J_i
\sqJ (R_{i\J})^\J\!\!-\lambda^\J_0 \sqdotJ=0 \ ,
\end{equation}
clearly verified by the $\sqdotJ$'s given in Eqs.\ (\ref{sqdotJD}) and (\ref{DJ}).

If two subsystems $A$ and $B$ are
non-interacting but in correlated states, the reduced state
operators obey the equations
\begin{mathletters}\label{separated}\begin{eqnarray}
\ddt{\rho_A}&=&-\frac{i}{\hbar}[H_A,\rho_A]\nonumber\\ &&-\sum_{\stackrel{\scriptstyle
J=1}{J\in A}}^M \frac{1}{ 2\tau_\J(\rho)} \big[\sqJ D_\J\!+\!D_\J^\dagger\sqJ\big]\Otimes (\rho_A)_\Jbar
\ , \\
 \ddt{\rho_B}&=&-\frac{i}{\hbar}[H_B,\rho_B] \nonumber\\ &&-\sum_{\stackrel{\scriptstyle
J=1}{J\in B}}^M \frac{1}{ 2\tau_\J(\rho)} \big[\sqJ D_\J\!+\!D_\J^\dagger\sqJ\big]\Otimes (\rho_B)_\Jbar
\ ,
\end{eqnarray}\end{mathletters}
where $(\rho_A)_\Jbar=\Tr_\J(\rho_A)$ and
$(\rho_B)_\Jbar=\Tr_\J(\rho_B)$.

The strong separability conditions (\ref{separability}) are
satisfied provided the internal-relaxation-time functionals are
either all constants or satisfy the following set of nontrivial
conditions.
\begin{mathletters}\label{tauseparability}\begin{eqnarray}
&&\tau_\J(\rho_A\Otimes\rho_B, H_A\Otimes I_B\!+\!I_A\Otimes H_B,
G_{iA}\Otimes I_B\!+\!I_A\Otimes G_{iB})\nonumber\\
&&\qquad=\left\{\begin{array}{ll} \tau_\J(\rho_A, \! H_A, \!
G_{iA})&\mbox{for $J\in A$}\\ \tau_\J(\rho_B, \! H_B, \!
G_{iB})&\mbox{for $J\in B$}\end{array}\right.\ , \\
&&\tau_\J(\rho, H_A\Otimes I_B\!+\!I_A\Otimes H_B, G_{iA}\Otimes
I_B\!+\!I_A\Otimes G_{iB})\nonumber\\
&&\qquad=\left\{\begin{array}{ll} {\tau_\J}' (\rho, \! H_A, \!
G_{iA})&\mbox{for $J\in A$}\\ {\tau_\J}' (\rho, \! H_B, \!
G_{iB})&\mbox{for $J\in B$}\end{array}\right. \quad {\rm for\
any}\ \rho \ .
\end{eqnarray}\end{mathletters}
It is noteworthy that Conditions \ref{tauseparability} are
satisfied by the maximal-entropy-production-rate ansatz, i.e., if
we assume that each $\tau_\J(\rho)$ is given by Eq.\
(\ref{taumin1J}) with strict equality.  Indeed, if $A$ and $B$ are
noninteracting, the structure [Eq.\ (\ref{DJ}c)] of each operator
$D_\J$ with $J\in A$  is such that any dependence on $H_B$ cancels
out,  moreover $\cov{H}{H}^\J=\cov{H_A}{H_A}^\J$. Thus, any
dependence on $H_B$ cancels out in Eq.\ (\ref{separated}a) and,
similarly, any dependence on $H_A$ cancels out in Eq.\
(\ref{separated}b).

It is interesting, however, that ${\rm d}\rho_A/{\rm d}t $ in
general may depend not only on the ``local" (reduced) state
operator $\rho_A$ but also on the overall state $\rho$ through the
operators $(B\ln\rho)^\J$ (with $J\in A$), thus determining a
collective-behavior effect on the local dynamics originating from
existing correlations. In fact, operators $D_\J$ are in terms of
$(\Delta S)^\J$  which may differ, when the subsystems are
correlated, from  operators $(\Delta S_A)^\J$. In other words, the
lack of interactions between two subsystems does guarantee that
the energy of each subsystem is conserved [Eqs.\
(\ref{separabilityenergy})] and that the reduced dynamics of each
subsystem is  local, in the sense of independent of the
Hamiltonian operator (and the other generators) of the other
subsystem, but does not necessarily imply that each subsystem
evolves independently of existing correlations with the other
subsystem.

Regarding the time evolution of correlations, in Ref.\
\cite{Cimento} we defined the correlation functional between two
subsystems $A$ and $B$,
\begin{equation}\label{correlation}
\sigma_{AB}(\rho)=\Tr(\rho\ln\rho) - \Tr_A(\rho_A\ln\rho_A) - \Tr_B(\rho_B\ln\rho_B)\ ,
\end{equation}
which is nonnnegative definite in general, and zero only if $\rho=\rho_A\Otimes \rho_B$.  The rate of change of the correlation can be written as
\begin{equation}\label{correlationrate}
\ddt{\sigma_{AB}(\rho)}=\dot\sigma_{AB}|_H- \dot\sigma_{AB}|_D\ .
\end{equation}
Based on our understanding of the equation of motion, we conjectured that $\dot\sigma_{AB}|_D $ should always be nonnegative, if it is true that the dissipative term can only destroy correlations between subsystems, but cannot create them.  However, this conjecture  remains to be proved. The corresponding entropy inequality to my knowledge has not yet  been studied.

\section{CONCLUSIONS}

I reviewed most previous results in Refs.\ \cite{thesis,Cimento,Beretta,Attractor,Fluorescence,Onsager} on the well-behaved nonlinear equation of motion I proposed for quantum thermodynamics that entails the second law and nonequilibrium steepest-entropy-ascent dynamics with Onsager's reciprocity  and Callen's fluctuation-dissipation relations. Together with the variational principle formulation derived in Ref.\ \cite{Gheorghiu} and the observation in Refs.\ \cite{HG,Gyft} that only the functional  $-\Boltz \Tr(\rho\ln\rho)$ can represent the physical entropy, we may conclude that all the results so far confirm that my equation of motion has all the necessary features to provide a self-consistent and conceptually-sound resolution of the century-old dilemma on the nature of entropy and irreversibility, alternative to Boltzmann's statistical approach and valid also for systems with few degrees of freedom.

 The nonlinear dynamics encompasses within a unified framework all the successful results of quantum mechanics, equilibrium and nonequilibrium thermodynamics.  It also holds the promise to provide a fundamental framework within which to address the currently unexplained evidence on loss of quantum coherence, to design new fundamental experiments, to examine new applications on the lines of those developed in Refs.\ \cite{Fluorescence,Korsch,Gheorghiu}, and to further investigate the dependence of the internal-relaxation-time functional on the state operator and physical constants, as well as possibly verify the ansatz proposed in this paper by which each indivisible subsystem follows the direction of steepest perceived entropy ascent at the highest rate compatible with the time--energy uncertainty principle.

The equation of motion satisfies the set of conditions discussed
in Appendix \ref{criteria} and, therefore, preserves most of the
traditional conceptual keystones of physical thought, including a
strongest form of the non-relativistic principle of causality, by
which future states of a strictly isolated system should unfold
deterministically from initial states along smooth unique
trajectories in state domain defined for all times (future as well
as past \cite{Proofs}).  Interestingly, while the maximum
entropy states are attractors in forward time, the
unitary-solutions boundary limit cycles of standard quantum mechanics are attractors in backward
time.

As pointed out by Onsager and Machlup \cite{Callen}, the fluctuation-dissipation relations cannot be derived in any rigorous way from the traditional Hamiltonian dynamical principles, unless these are complemented by some additional postulate closely related to the additional principles, assumptions, or approximations needed to derive the Onsager reciprocity relations. This is sometimes referred to as the irreversibility paradox. In other words, in order to infer any feature of irreversibility (including its very existence) from the irreducibly reversible dynamical principle of standard Hamiltonian mechanics, we must complement it with some additional postulate that seems to contradict it.

 Within our nonlinear quantum (thermo)dynamics based on Eqs. (\ref{EquationComposite}) and (\ref{EquationSingle}) the paradox is resolved. The augmented state domain ansatz broadens the set of conceivable states but includes the standard pure states, and the nonlinear equation of motion describes irreversibile time evolutions and entails reciprocity and fluctuation-dissipation relations, but maintains the standard unitary dynamics of pure states.

Finally, from the Heisenberg time-energy uncertainty principle I derived a lower bound for the internal-relaxation-time functionals from which follows an upper bound for the rate of entropy production.  Consequently, I proposed a physically intriguing maximal-entropy-production-rate ansatz, by which each indivisible subsystem follows the direction of steepest perceived entropy ascent at the highest rate compatible with the time--energy uncertainty principle. If this ansatz is experimentally verified, the nonlinear dynamics is complete and self-consistent, with no need of new physical constants.

\section*{Acknowledgements}

Work supported in part by grants from the
italian Ministero dell'Istruzione, dell'Universit\`a e della Ricerca
(MIUR) and the italian Istituto Nazionale di Fisica della Materia (INFM).

\appendix
\section{CRITERIA FOR A GENERAL (NONLINEAR)
QUANTUM DYNAMICS COMPATIBLE WITH THERMODYNAMICS}\label{criteria}

Within a quantum theory that accepts the augmented set of true quantum states described by state operators $\rho$ without the restriction $\rho^2=\rho$, and a nonlinear dynamical law for a strictly isolated system, the following demanding set of conditions should be satisfied in order for the theory to be compatible or, better, imply the second law of thermodynamics without contradicting the fundamental results of standard quantum mechanics (QM).  Obviuosly these are the criteria I followed in designing Eqs.\ (\ref{EquationComposite}) and (\ref{EquationSingle}), and discussed at length in Refs.\ \cite{Beretta,Taormina}.

Conditions 6, 7 and 8 are closely related to the condition recently referred to as strong separability \cite{Czachor}.  I also added a condition on correlation and entanglement to reflect the need to avoid, and possibly resolve, physical inconsistencies related to nonlocality issues, as well as a strong causality condition that is nontrivial and quite demanding both from the conceptual and the technical mathematical points of view.

Certainly, when viewed from different perspectives --- e.g., different physical interpretations of the augmented state domain $\rho^2\ne\rho$ ansatz, of the role of the nonlinear extension of the Schr\"odinger equation of motion, of the Shannon-von Neumann entropy functional $-\Boltz\Tr(\rho\ln\rho)$ versus other nonextensive functionals, of the role of the system's environment and the measuring apparati, and so on --- some authors might view this set of conditions as too strong in many respects.  Nevertheless, my equation of motion demonstrates that at least a satisfactory dynamics exists which satisfies all such conditions and, in my view, features  a number of intriguing, unifying and far-reaching implications.

\subsection*{1.\ Causality. Forward and backward in time}
Considering the set $\Prho$ of all linear, hermitian, nonnegative-definite, unit-trace  operators $\rho$ on $\Hil$, every solution of the equation of motion, i.e., every trajectory $u(t,\rho)$ which at time $t=0$ passes through state $\rho$ in $\Prho$, should lie entirely in $\Prho$ for all times $t$, $-\infty<t<+\infty$.

\subsection*{ 2.\ Standard QM unitary evolution of $\rho^2=\rho$ states}
The unitary time evolution of the states of QM according to the Schr\"odinger
equation of motion must be compatible with the more general dynamical law.  These trajectories, passing through any state
$\rho$ such that $\rho^2=\rho$ and entirely contained in the state domain of
Quantum Mechanics, must be solutions also of the more general dynamical law. In view of the fact that the states of
QM are the extreme points of the augmented state domain, the
trajectories of QM must be boundary solutions (limit cycles) of the dynamical
law.

If the complete dynamics preserves the feature of uniqueness of
solutions throughout the augmented state domain, then pure states
can only evolve according to the Schr\"odinger equation of motion.
In this case, no trajectory can enter or leave the state domain of
QM and by continuity, there must be trajectories that approach
indefinitely these boundary solutions (of course, this can only
happen backward in time, as $t\to -\infty$).

\subsection*{ 3.\ Conservation of energy and number of particles }
If the system is isolated, the value of the energy functional $e(\rho)=\Tr(\rho H)$, where $H$ is the standard QM Hamiltonian operator,
must remain invariant along every trajectory.  If the isolated system consists
of a variable amount of a single type of particle with a number operator $N$
that commutes with $H$, then also the value of the
number-of-particle functional $n(\rho)=\Tr(\rho N)$ must remain invariant along every
trajectory.  If the isolated system consists of $r$ types of particles each with
variable amount and each with a number operator $N_i$ that commutes with the
Hamiltonian $H$, then also the value of each number-of-particle functional $n_i(\rho)=\Tr(\rho N_i)$ must remain invariant along every trajectory.  Depending on the type of system, there may be other time-invariant functionals.

\subsection*{ 4.\ Stability of the thermodynamic equilibrium states. Second law}
A state operator $\rho$ represents an equilibrium state if ${\rm
d}\rho/{\rm d}t=0$ when the system is isolated [e.g., $H\ne H(t)$].
For each given set of feasible values of the energy functional
$e(\rho)$ and the number-of-particle functionals $n_i(\rho)$ (i.e., the
functionals that must remain invariant according to Condition 2 above),
among all the equilibrium states that the dynamical law may admit there
must be one and only one which is globally stable (definition and
discussion in Appendix \ref{stability}).  This stable equilibrium state
must represent the corresponding state of equilibrium thermodynamics
and, therefore, must be of the form given by Eq.\ (\ref{stableeq}). All
the other equilibrium states that the dynamical law may admit must not
be globally stable.

\subsection*{5.\ Entropy nondecrease. Irreversibility}
The principle of nondecrease of entropy must be satisfied, i.e., the rate of
change of the entropy functional $-\Boltz\Tr(\rho\ln\rho)$ must be nonnegative along every trajectory.

\subsection*{ 6.\ Non-interacting subsystems. Separate energy conservation}
For an isolated system composed of two distinguishable subsystems $A$ and $B$ with associated
Hilbert spaces $\Hil^A$ and $\Hil^B$, so that the Hilbert space of the system is
$\Hil=\Hil^A\Otimes\Hil^B$, if the two subsystems are non-interacting, i.e., the
Hamiltonian operator $H = H_A\Otimes I_B + I_A\Otimes H_B$, then the functionals
$\Tr[(H_A\Otimes I_B)\rho]= \Tr_A(H_A\rho_A) $ and $\Tr[(I_A\Otimes H_B)\rho]= \Tr_B(H_B\rho_B)$ represent the
energies of the two subsystems and must remain invariant along every trajectory, even if the states of $A$ and $B$ are correlated, i.e., even if $\rho\ne \rho_A\Otimes\rho_B$. Of course, $\rho_A=\Tr_B(\rho)$, $\rho_B=\Tr_A(\rho)$, $\Tr_B$ denotes the partial trace over $\Hil^B$ and $\Tr_A$
the partial trace over $\Hil^A$.

\subsection*{ 7.\ Independent states. Weak separability. Separate entropy nondecrease }
Two distinguishable subsystems $A$ and $B$ are in independent
states if the state operator $\rho=\rho_A\Otimes\rho_B$, so that
the entropy operator $S =-\Boltz\ln\rho= S_A\Otimes I_B +
I_A\Otimes S_B= -\Boltz [B_A\ln\rho_A\Otimes I_B + I_A\Otimes
B_B\ln\rho_B ]$. For permanently non-interacting subsystems, every
trajectory passing through a state in which the subsystems are in
independent states must maintain the subsystems in independent
states along the entire trajectory.  When two uncorrelated systems
do not interact with each other, each must evolve in time
independently of the other.

In addition, if at some instant of time two subsystems $A$ and
$B$,  not necessarily non-interacting, are in independent states,
then the instantaneous rates of change of the subsystem's entropy
functionals $-\Boltz\Tr(\rho_A\ln\rho_A)$ and
$-\Boltz\Tr(\rho_B\ln\rho_B)$ must both be nondecreasing in time.

\subsection*{8.\ Correlations, entanglement and locality. Strong separability}
Two non-interacting subsystems $A$ and $B$ initially in correlated
states  (possibly due to a previous interaction that has then been
turned off) should each proceed in time towards less correlated
states or, at least, maintain the same level of quantum
entanglement.  The generation of quantum entanglement between
interacting subsystems should emerge only through the
Schr\"odinger-von Neumann term $-i[H,\rho]/\hbar$ of the equation
of motion, whereas the other terms, that might entail loss of
correlations between subsystems, must not be able to create them.
This condition is perhaps too strong and its validity for my
equation is still only conjectural. In any case, the dynamics
should not generate locality problems, i.e., faster-than-light
communication between noninteracting subsystems, even if in
entangled or correlated states.  In other words, when subsystem
$A$ is not interacting with subsystem $B$, it should never be
possible to influence the local observables of $A$ by acting only
on the interactions within $B$, such as switching on and off
parameters or measurement devices within $B$. This does not mean
that existing correlations between $A$ and $B$ established by past
interactions should have no influence whatsoever on the time
evolution of the local observables of either $A$ or $B$. In
particular, I see no physical reason to request that two different
states $\rho$ and $\rho'$ such that $\rho'_A=\rho_A$ should evolve
in such a way that ${\rm d}\rho'_A /{\rm d}t ={\rm d}\rho_A /{\rm
d}t$ whenever $A$ is isolated (but not uncorrelated) from the rest
of the overall system. For example, state $\rho'$ could be the
maximum entropy stable equilibrium state (and, therefore,
$\rho'=\rho'_A\Otimes\rho'_B$, ${\rm d}\rho' /{\rm d}t =0$)
whereas in state $\rho$ subsystems $A$ and $B$ could be correlated
and evolving in time towards the stable equilibrium state or
$\rho$ could even be a pure entangled state evolving along a
unitary trajectory according to Condition 2 above, and therefore
it would never reach stable equilibrium.

\section{LYAPUNOV STABILITY AND THERMODYNAMIC
STABILITY}\label{stability}

The condition concerning the stability of the thermodynamic equilibrium states
is extremely restrictive and requires further discussion.

In order to implement Condition 5 in Appendix \ref{criteria}, we
need to establish the relation between the notion of stability
implied by the second law of thermodynamics \cite{Lyapunov,Books}
and the mathematical concept of stability.  An equilibrium state
is stable, in the sense required by the second law, if it can be
altered to a different state only by interactions that leave net
effects in the state of the enviromment.  We call this notion of
stability {\it global stability}.  The notion of stability
according to Lyapunov is called {\it local stability}.

We denote the trajectories generated by the dynamical law on our state domain by
$u(t,\rho)$, i.e., $u(t,\rho)$ denotes the state at time $t$ along the
trajectory that at time $t=0$ passes through state $\rho$.  A state $\rho_e$ is
an equilibrium state if and only if $u(t,\rho_e)=\rho_e$ for all times $t$.  An
equilibrium state $\rho_e$ is {\it locally stable (according to
Lyapunov)} if and only if for every $\epsilon>0$ there is a
$\delta(\epsilon)>0$ such that $d(\rho,\rho_e)<\delta(\epsilon)$ implies
$d(u(t,\rho),\rho_e)<\epsilon$ for all $t>0$ and every $\rho$, i.e., such that
every trajectory that passes within the distance $\delta(\epsilon)$ from state
$\rho_e$ proceeds in time without ever exceeding the distance $\epsilon$ from
$\rho_e$.  Conversely, an equilibrium state $\rho_e$ is unstable if
and only if it is not locally stable, i.e., there is an $\epsilon>0$ such that
for every $\delta>0$ there is a trajectory passing within distance $\delta$ from
$\rho_e$ and reaching at some later time farther than the distance $\epsilon$
from $\rho_e$.

The Lyapunov concept of instability of equilibrium is clearly equivalent to that
of instability stated in thermodynamics according to which an equilibrium state
is unstable if, upon experiencing a minute and short lived influence by some
system in the environment (i.e., just enough to take it from state $\rho_e$ to a
neighbouring state at infinitesimal distance $\delta$), proceeds from then on
spontaneously to a sequence of entirely different states (i.e., farther than
some finite distance $\epsilon$).

It follows that the concept of stability in thermodynamics implies that of
Lyapunov local stability.  However, it is stronger because it also excludes the
concept of {\it metastability}.  Namely, the states of equilibrium
thermodynamics are {\it global} stable equilibrium states in the sense that not
only they are locally stable but they cannot be altered to entirely different
states even by means of interactions which leave temporary but finite effects in
the environment.  Mathematically, the concept of metastability can be defined as
follows.  An equilibrium state $\rho_e$ is {\it metastable} if and only if it
is locally stable but there is an  $\eta>0$ and an $\epsilon>0$ such that for
every $\delta>0$ there is a trajectory $u(t,\rho)$ passing at $t = 0$ between
distance $\eta$ and $\eta+\delta$ from $\rho_e$, $\eta<d(u(0,\rho),\rho_e) <
\eta+\delta$, and reaching at some later time $t>0$ a distance farther than
$\eta+\epsilon$, $d(u(t,\rho),\rho_e) \ge \eta+\epsilon$.  Thus, the
concept of global stability implied by the second law is as follows.  An
equilibrium state $\rho_e$ is {\it globally stable} if for every $\eta>0$ and
every $\epsilon>0$ there is a $\delta(\epsilon,\eta)>0$ such that every
trajectory $u(t,\rho)$ with $\eta<d(u(0,\rho),\rho_e) <
\eta+\delta(\epsilon,\eta)$, i.e., passing at time $t = 0$ between distance
$\eta$ and $\eta+\delta$ from $\rho_e$, remains within $d(u(t,\rho),\rho_e) <
\eta+\epsilon$ for every $t>0$, i.e., proceeds in time without ever exceeding
the distance $\eta+\epsilon$.

The second law requires that for each set of values of the
invariants $\Tr(\rho H)$ and $\Tr(\rho G_i)$ (as many as required
by the structure of the system), and of the parameters embedded in
the Hilbert space $\Hil$ and the Hamiltonian $H$ describing the
external forces (such as the size of a container), there is one
and only one globally stable equilibrium state. Thus, the
dynamical law may admit many equilibrium states that all share the
same values of the invariants and the parameters, but among all
these only one is globally stable, i.e., all the other equilibrium
states are either unstable or metastable.

Interestingly, we may use this condition to show that a unitary
(Hamiltonian) dynamical law would be inconsistent with the
second-law stability requirement. A unitary dynamical law in the
augmented kinematics would be expressed by an equation of motion
$i\hbar \dot\rho=[H,\rho]$  with trajectories $u(t,\rho) =
U(t)\rho U^{-1}(t)$ with $U(t) = \exp(-itH/\hbar)$.  Such a
dynamical law would admit as equilibrium states all the states
$\rho_e$ such that $\rho_eH=H\rho_e$.  Of these states there are
more than just one for each set of values of the invariants.  With
respect to the metric $d(\rho_1,\rho_2) = \Tr|\rho_1-\rho_2|$, it
is easy to show that every trajectory $u(t,\rho)$ would be
equidistant from any given equilibrium state $\rho_e$, i.e.,
$d(u(t,\rho),\rho_e) = d(u(0,\rho),\rho_e)$ for all $t$ and all
$\rho$.  Therefore, all the equilibrium states would be globally
stable and there would be more than just one for each set of
values of the invariants, thus violating the second-law
requirement.

The entropy functional $-\Boltz\Tr(\rho\ln\rho)$  plays a useful
role in proving the stability of the states of equilibrium
thermodynamics [Eq.\ (\ref{stableeq})] provided that the dynamical law
guarantees that $-\Boltz\Tr [u(t,\rho)\ln u(t,\rho)]\ge
-\Boltz\Tr(\rho\ln\rho)$ for every trajectory, i.e., provided
Condition 6 above  is satisfied. The proof of this is nontrivial
and is given in Ref.\ \cite{Lyapunov} where,  however, we also
show that the entropy functional is not a Lyapunov function, even
if, in a strict sense that depends on the continuity and the
conditional stability of the states of equilibrium thermodynamics,
it does provide a criterion for the  stability of these states.
Anyway, even if the entropy were a Lyapunov function,  this would
suffice only to guarantee the local stability of the states of
equilibrium thermodynamics but not to guarantee, as required by
the second law,  the instability or metastability of all the other
equilibrium states.

\section{CHARACTERISTIC TIMES}\label{times}

Using Eq.\ (\ref{rhodot}), the rate of change of the mean functional
$f(\rho)=\Tr(\rho F)$ may be written as
\begin{equation}
\ddt{f(\rho)}=\Tr\left(\ddt{\rho}F\right)=2\left(\sq
F\left|E\right.\right)\ .
\end{equation}

For the Schr\"odinger--von Neumann evolution, the characteristic time of change of $f(\rho) $ may be defined as \cite{Messiah}
\begin{equation}
\frac{1}{\tau^2_{FH}}=\frac{[{\rm d}f(\rho)/{\rm d}t]^2}{\cov{F}{F}}=\frac{4\left(\sq \Delta F\left|\sqdotH\right.\right)^2}{\cov{F}{F}} \ .
\end{equation}

Because operators $\sq \Delta F/\sqrt{\cov{F}{F}}$ are unit norm, in the sense that $(\sq \Delta F|\sq \Delta F)/\cov{F}{F}=1$, it follows that the characteristic times $\tau_{FH}$ are bounded by the value attained for an operator $F$ such that $\sq \Delta F$ is in the same direction as $\sqdotH$, i.e., such that
\begin{equation}
\frac{\sq \Delta F}{\sqrt{\cov{F}{F}}}=\frac{\sqdotH}{ \sqrt{\left(\sqdotH\left|\sqdotH\right.\right)}} \ .
\end{equation}
Therefore,
\begin{equation}
\frac{1}{\tau^2_{FH}}\le 4 \left(\sqdotH\left|\sqdotH\right.\right)= \frac{1}{\tau^2_{H}} \ .
\end{equation}
For this reason, in Eq.\ (\ref{normH1}) we take $\tau_H$ equal to the lower bound of the $\tau_{FH}$'s.

By analogy, but considering the understanding of the contribution of each subsystem to the overall-system dynamics embedded in our nonlinear dynamics, we define the characteristic time of the dissipative change of the mean functional $f(\rho)=\Tr(\rho F)$ due to the $J$-th constituent subsystem as
\begin{equation}
\frac{1}{\tau^2_{FD\J}}=\frac{[{\rm D}f(\rho)/{\rm D}t]^2}{\cov{F}{F}^\J}=\frac{4\left(\sqJ (\Delta F)^\J\left|\sqdotJD\right.\right)^2}{\cov{F}{F}^\J} \ .
\end{equation}

Again, because operators $\sqJ (\Delta F)^\J/\sqrt{\cov{F}{F}^\J}$ are unit norm, in the sense that $(\sqJ (\Delta F)^\J |\sqJ (\Delta F)^\J)/\cov{F}{F}^\J=1$, it follows that the characteristic times $\tau_{FD\J}$ are bounded by the value attained for an operator $F$ such that $\sqJ (\Delta F)^\J $ is in the same direction as $\sqdotJD$, i.e., such that
\begin{equation}
\frac{\sqJ (\Delta F)^\J }{\sqrt{\cov{F}{F}^\J}}=\frac{\sqdotJD}{ \sqrt{\left(\sqdotJD\left|\sqdotJD\right.\right)}} \ .
\end{equation}
Therefore,
\begin{equation}
\frac{1}{\tau^2_{FD\J}}\le 4 \left(\sqdotJD\left|\sqdotJD\right.\right) =\frac{1}{\tau^2_{D\J}}\ .
\end{equation}
For this reason, in Eqs.\ (\ref{normD}) and (\ref{normDJ}) we take $\tau_D$ and $ \tau_{D\J}$ equal to the respective lower bounds of the $\tau_{FD\J}$'s.

\section{Special form of the equation of motion}\label{special}

If, for a given state operator $\rho$, we construct the set $\{\sq X_{j } \}$ so as to be an orthogonal extension of the orthonormal subset $\{\sq A_{i } \}$, i.e., with  $X_i=A_i$ for $i\le a$ and $(\sq X_i|\sq X_j)=\delta_{ij}$ for all $i$ and $j$, then $D=-\sum_{j>a}f_j\,\sq X_j$ [Eq.\ (\ref{DAf})] and the one-constituent equation of motion reduces to the (only apparently linear) form
\begin{equation}\label{apparentlinear}
\ddt{\rho}=-\frac{i}{\hbar}[H,\rho]+\frac{1}{2\tau(\rho)}\sum_{j>a} f_j(\rho)\,\{X_j(\rho),\rho\}\ ,
\end{equation}
where the dependences of $f_j$, $X_j$ and $\tau$ on $\rho$ are evidenced in order to emphasize the nonlinearity.

For such special choice of the $X_j$'s, assuming $X_1=I$, we have $x_1(\rho)=\Tr(\rho)=1$, $x_{i\ne 1}(\rho)=\Tr(\rho X_{i\ne 1})=0$, $\cov{X_1}{X_i}=0$, $\cov{X_{i\ne 1}}{X_{j\ne 1}}=\delta_{ij}$, $L_{ij}(\rho)=0$ and ${\rm D}x_i(\rho)/ {\rm D}t=0$ for $i$ or $j\le a$, $L_{ij}(\rho)=\delta_{ij}/\tau(\rho)$ and ${\rm D}x_i(\rho)/ {\rm D}t=f_i(\rho)/\tau(\rho)$ for $i$ and $j> a$, so that
\begin{eqnarray}
\ddt{s(\rho)}&=&\frac{\Boltz}{\tau(\rho)}\sum_{k>a} f_k(\rho)^2 \ , \\
\tau(\rho)|_{\rm min}&=&\frac{\hbar}{2\cov{H}{H}}\sqrt{\sum_{k>a} f_k(\rho)^2}\ , \\
\left. \ddt{s(\rho)}\right|_{\rm max}&=&\frac{2\Boltz}{\hbar}\sqrt{\cov{H}{H}}\sqrt{\sum_{k>a} f_k(\rho)^2} \ .
\end{eqnarray}
where in the last two equations we made use of Eqs.\ (\ref{rateSfinal}) and Relations (\ref{taumin1}) and (\ref{taumin2}) with strict equality.

In view of the dependence of the $X_j$'s on $\rho$ and, therefore, on time, the apparently simple form of Eq.\ (\ref{apparentlinear}) may not be as useful as it seems.

\section{Extension to time-varying rates of the generators of the
motion}\label{ExtensionRates}

For the purposes of quantum thermodynamics, in my view the dissipative part of the equation of motion should not account for rates of change of the mean values of the generators of the motion other than through the Hamiltonian term, consistently with all the results of standard QM.

Nevertheless, the mathematical extension of Eqs.\ (\ref{EquationComposite}) and (\ref{EquationSingle}) to (artificially) imposed rates $\dot r_i \ne 0$ (arbitrarily specified as functions of time) is straightforward and may be useful in applications or other frameworks \cite{ASME}.

It does emerge naturally from the maximization problem (\ref{maxprob}); it does so, however, implicitly, through substitution of Eq.\ (\ref{maxS}) back into the constraint Equations (\ref{maxprob}b). Instead, the explicit form in terms of projections and, therefore, the equivalent expressions by means of Gram determinants amount to assuming, for the operator $D$ in the equation of motion, instead of Eq.\ ({\ref{EquationSingle}c) or the equivalent Eqs.\ (\ref{Dgram}),
\begin{eqnarray}
D &= &[\sq\ln\rho]_{\bot\Ell {\{\sq R_{i } \}}}\nonumber\\
& &+\sum_j \alpha_j [\sq R_j]_{\bot\Ell {\{\sq\ln\rho ,\sq R_{i\ne j } \}}}\ ,
\end{eqnarray}
where the explicit expression of the $j$-th term in the summation is given by
Eq.\ (\ref{Dgram}b) where we interchange everywhere $\sq\ln\rho$ with $\sq R_j$ and
\begin{equation}
\alpha_j=\tau(\rho)\, \dot r_j\, \frac{\Gamma(\sq\ln\rho,\{\sq R_{i\ne j}\})}{\Gamma(\sq\ln\rho,\{\sq R_i\})}\ .
\end{equation}
Geometrically, the additional terms are in the steepest-$r_j$-ascent direction compatible with maintaining constant the mean values of the other generators and the entropy.

Analogous obvious extension to the composite system case can be obtained by assuming for the operator $D_J$, instead of Eq.\ (\ref{EquationComposite}),
\begin{eqnarray}
D_J &=& [\sqJ(B\ln\rho)^\J]_{\bot\Ell {\{\sqJ (R_{i })^\J \}}}\nonumber\\
&+&\sum_j \alpha^J_j [\sqJ (R_j)^\J]_{\bot\Ell {\{\sqJ(B\ln\rho)^\J ,\sqJ (R_{i\ne j })^\J \}}}\ ,
\end{eqnarray}
with the $\alpha^\J_j$'s such that
\begin{equation}
\dot r_j =\sum_{J=1}^M \frac{\alpha^\J_j }{\tau_\J (\rho)} \frac{\Gamma(\sqJ(B\ln\rho)^\J,\{\sqJ( R_i)^\J\})}{\Gamma(\sqJ(B\ln\rho)^\J,\{\sqJ( R_{i\ne j})^\J\})} \ .
\end{equation}

\section{Extension to other entropy or mean value
functionals}\label{ExtensionEntropy}

For the purposes of quantum thermodynamics, in my view
the necessary entropy functional is $-\Boltz\Tr(\rho\ln\rho)$, for
the reasons in Refs.\
\cite{Cimento,Beretta,Onsager,Taormina,HG,Gyft}.

However, in view of  the recent literature on nonextensive quantum
theories, as suggested also in Ref.\ \cite{Gheorghiu}, it may be
useful to note that the entire formalism of my equation of motion
can be readily reformulated in the case of any other well-behaved
entropy functional $s(\rho)$ and set of nonlinear generator
functionals $r_i(\rho)$.

For the single constituent  system it suffices to substitute throughout
the operator $(-2\Boltz\sq I+) \sq S=(-2\Boltz\sq)-2\Boltz\sq \ln\rho$
(notice that in the Gram determinants the addenda in parantheses cancel
out) with the new entropy gradient operator, $\hpardsq{s(\rho)}$, and
the operators $2 \sq R_i$ with [operators which when symmetrized
($\{A,A^\dagger\}/2$) are equal to] the gradient operators of the new
generator functionals, $\hpardsq{r_i(\rho)}$.

However, for a composite  system, consistently with the
nonextensivity of these theories, it may be difficult or not at
all possible to identify the operators corresponding to $(S)^\J$
and $(R_i)^\J$ representing the subsystems' local perceptions of
the entropy and the generators.

\section{A noteworthy equation for a composite system that fails to meet a separability condition}\label{wrong}

It is interesting to note that the role of the $\sqJ$ operators in Eq.\ (\ref{EquationComposite}) [and that of $\sq$ in Eq.\ (\ref{EquationSingle})] is formally useful but only auxiliary, because wherever there is a $\sqJ$, another $\sqJ$ comes in front or behind it.  In fact, in my doctoral thesis \cite{thesis} the equation of motion was in terms of $\rho_\J$ only.  I realized the usefulness of the $\sqJ$ formalism only later, in connection with the proof of the steepest-entropy-ascent geometric property \cite{Beretta}.

By allowing a more relevant role of the $\sq$ operator, an alternative to the construction in Eq.\ (\ref{rhodotM}) may appear formally better, and suggest the apparently alternative equation of motion based on the following definitions
\begin{mathletters}\label{wrongEquation}\begin{eqnarray}
&\sqdotD =-\sum_{J=1}^M \frac{1}{2\tau_\J(\rho)}D_\J' \Otimes \sqJbar \ ,& \\
\label{privateFwrong}&(\sq F)^\J_\sqempty =\Tr_\Jbar [(I_\J\Otimes \sqJbar) \sq F]\ ,& \\
&D_\J'= [ (\sq\ln\rho)^\J_\sqempty ]_{\bot\Ell {\{(\sq)^\J_\sqempty ,(\sq H)^\J_\sqempty , [(\sq G_{i })^J_\sqempty ]\}}} \ .&
\end{eqnarray}\end{mathletters}
In fact, it can be readily verified that for a given $F$ on $\Hil$ with $[F,H]=0$ the rate of change of $\Tr(\rho F)$ would be zero if and only if $(\sq F)^\J_\sqempty $ is in
$\Ell {\{(\sq)^\J_\sqempty ,(\sq H)^\J_\sqempty , [(\sq G_{i })^J_\sqempty ]\}}$, and so $\Tr (\rho)$, $\Tr (\rho H)$, and $\Tr(\rho G_i)$ would be conserved.  The expressions for the rate of entropy production, the Onsager relations and the other results would be almost identical to those obtained from my equation, except for the substitution throughout of $\sqJ (F)^\J$ with $(\sq F)^\J_\sqempty $.

However, the resulting dynamics would fail to satisfy at least the important property expressed by Eq.\ (\ref{separabilityenergy}), because it  can be verified that if $H=H_\J\Otimes I_\Jbar+I_\J\Otimes H_\Jbar$ but $\rho\ne\rho_J\Otimes \rho_\Jbar$ then separate conservation of the non-interacting subsystem's energy would not be guaranteed.


\begin{references}
\bibitem{NotaSquareRoot}The $\rho=\gamma\gamma^\dagger$ formalism
proposed by Reznik \cite{Reznik} and adopted by Gheorghiu-Svirschevski
\cite{Gheorghiu}, extends my $\rho=(\sq)^2$ formalism but, as shown in the present paper, is not necessary to derive my equation.  Conceptually, it is troublesome because it leaves unspecified the choice of $\gamma$ for a given $\rho$, and risks to imply the existence of yet unidentified physical observables that would render the state operator insufficient.  Formally, it may suggest the adoption of Eq.\ (\ref{wrongEquation}) for a composite system (properly extended to the complex square root formalism), which has many correct properties but fails to meet a necessary separability requirement (see Appendix \ref{wrong}).
\bibitem{Note}This notation includes the special cases in which subsystem $J$ consists of only, say, the 4-th type of particle (in which case $N_{i\J}$ is the null operator for every $i\ne 4$) possibly with a fixed amount, say, 5 particles (in which case $N_{4\J}=5 I_\J$). Note that some $N_{iJ}$ may be the null operator if subsystem $J$ does not contain particles of type $i$. Even in this case the set of generators \{$R_{i\J}$\} remains well-defined.
\bibitem{Proofs}For rigorous proofs, the reader is referred to the original papers, where only some technical general aspects of the mathematics of the nonlinear equation for $\dim\Hil=\infty$ were left unresolved or based on technical conjectures, especially those related to the global stability of the thermodynamic equilibrium states, the instability or metastability of the other equilibrium states and all boundary solutions and limit cycles, and the existence and uniqueness in the general case of all solutions both forward and backwards in time, i.e., for $-\infty<t<\infty$. For a two-level system, instead, all these proofs are complete \cite{Attractor}.  The work in Ref.\ \cite{Gheorghiu} provides a further proof of preservation of positivity.  That in Refs.\ \cite{Lyapunov,Hiai} completes the proofs regarding stability.
\end{references}
\end{document}